\documentclass[amsmath,amssymb,prr,reprint,superscriptaddress]{revtex4-2} 

\usepackage {graphicx}
\usepackage{dcolumn}
\usepackage{bm}
\usepackage[whole]{bxcjkjatype}
\usepackage{verbatim}
\usepackage{color}
\usepackage{physics}

\usepackage{caption}
\usepackage[position=top,singlelinecheck=false]{subcaption}

\newcommand{\ii}{\ensuremath{\mathrm{i}}}

\begin{document}

\title{
Hybrid quantum-classical algorithm for computing imaginary-time correlation functions
}

\author{Rihito Sakurai}
\email{sakurairihito@gmail.com}
\affiliation{Department of Physics, Saitama University, Saitama 338-8570, Japan}

\author{Wataru Mizukami}
\affiliation{Center for Quantum Information and Quantum Biology, Osaka University, Osaka 565-0871, Japan}
\affiliation{JST, PRESTO, 4-1-8 Honcho, Kawaguchi, Saitama 332-0012, Japan}

\author{Hiroshi Shinaoka}
\affiliation{Department of Physics, Saitama University, Saitama 338-8570, Japan}
\affiliation{JST, PRESTO, 4-1-8 Honcho, Kawaguchi, Saitama 332-0012, Japan}

\date{\today}

\begin{abstract}
Quantitative descriptions of strongly correlated materials pose a considerable challenge in condensed matter physics and chemistry. 
A promising approach to address this problem is quantum embedding methods.
In particular, the dynamical mean-field theory (DMFT) maps the original system to an effective quantum impurity model comprising correlated orbitals embedded in an electron bath.
The biggest bottleneck in DMFT calculations is numerically solving the quantum impurity model, 
i.e., computing the Green's function.
Past studies have proposed theoretical methods to compute the Green's function of a quantum impurity model in polynomial time using a quantum computer. 
So far, however, efficient methods for computing the imaginary-time Green's functions have not been established despite the advantages of the imaginary-time formulation.
We propose a quantum--classical hybrid algorithm for computing imaginary-time Green's functions on quantum devices with limited hardware resources by applying the variational quantum simulation. Using a quantum circuit simulator, we verified this algorithm by computing Green's functions for a dimer model as well as a four-site impurity model obtained by DMFT calculations of the single-band Hubbard model, although our method can be applied to general imaginary-time correlation functions. \end{abstract}

\maketitle

\section{Introduction}\label{sec:introduction}

The accurate and efficient computation of quantum many-body systems is critical in computational physics.
As Feynman pointed out~\cite{feynman2018simulating}, quantum computers, which use quantum mechanics as their computational principle, 
are expected to efficiently simulate quantum systems.
There are obvious advantages in this approach in the field of strongly correlated electron 
such as frustrated magnetic compounds and high-temperature superconductors, which are difficult to simulate using classical computers~\cite{lloyd1996universal, whitfield2011simulation, bauer2020quantum}.
However, even if a quantum computer with thousands or tens of thousands of logical qubits is realized, simulating macroscopic degrees of freedom in solids remains challenging.

\begin{figure}
    \vspace{-15mm} 
    \centering
    \begin{flushleft}
    \end{flushleft}
    \includegraphics[width=1.0\linewidth]{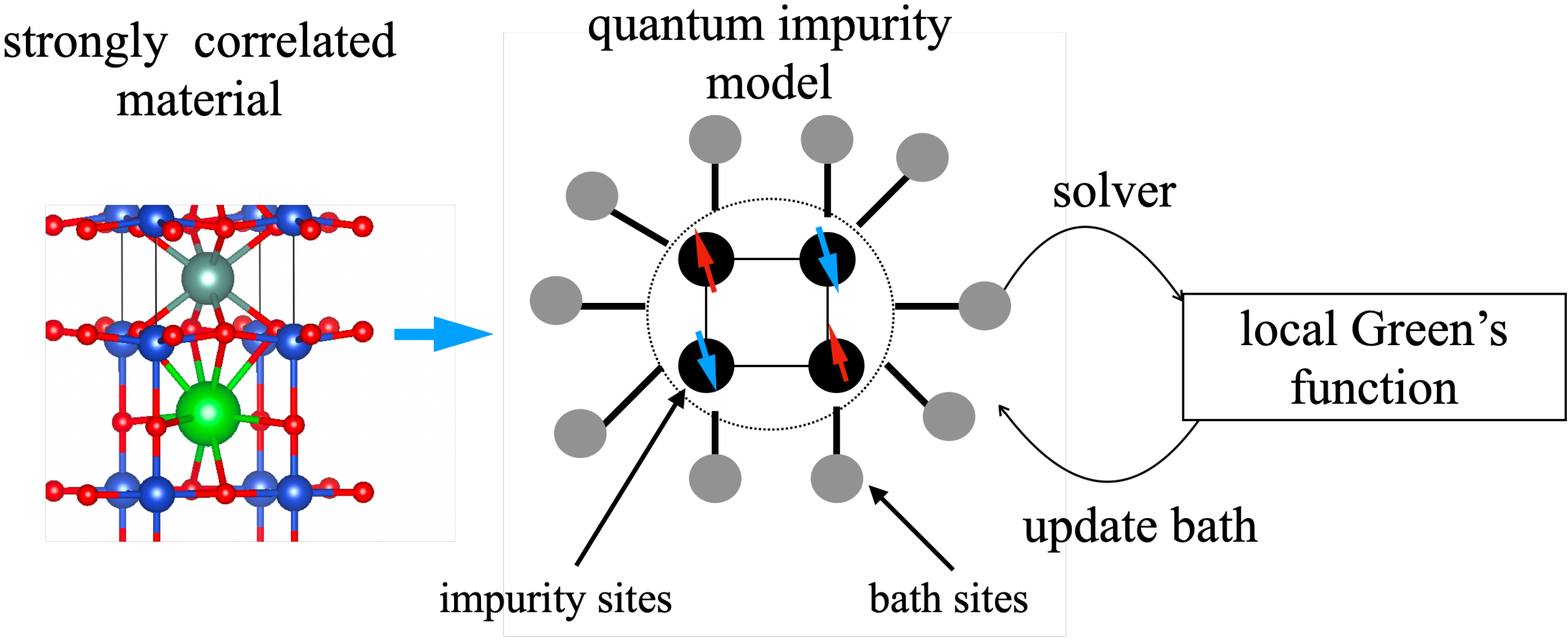}
    \vspace{-15mm} 
    \caption {
    Schematic of dynamical mean-field calculations and quantum impurity models.
    }
    \label{fig:dmft}
\end{figure}

We need a framework to reduce the degrees of freedom of solids to make them realistically computable by quantum computers.
One promising approach is the combination of quantum computing and quantum embedding theories such as the dynamical mean-field theory (DMFT)~\cite{georges1996dynamical,kotliar2006electronic}.
As illustrated in Fig.~\ref{fig:dmft}, DMFT maps the original system to an effective quantum impurity model comprising correlated orbitals embedded in an electron bath.
The biggest bottleneck in DMFT calculations is solving the quantum impurity model numerically, that is, computing the Green's function.
Although advanced classical algorithms such as the tensor network~\cite{PhysRevB.101.041101, wolf2015imaginary, bauernfeind2017fork} and quantum Monte Carlo~\cite{RevModPhys.83.349} have been developed, their applications are limited to a few correlated orbitals owing to the exponential growth of the computational cost with respect to the number of correlated orbitals.
This difficulty originates from the notorious negative sign problems and the rapid growth of quantum entanglement entropy.

It is highly desirable to solve the quantum impurity model with a quantum computer and overcome the above-mentioned difficulties.
In recent years, theoretical proposals have been made to compute the Green's function in polynomial time using a quantum computer~\cite{PhysRevA.92.062318,bauer2016hybrid,kreula2016non}.
However, these proposals assume a fault-tolerant quantum computer, which is beyond the capabilities of quantum devices with limited hardware resources such as noisy intermediate-scale quantum (NISQ) devices~\cite{preskill2018quantum}.
Therefore, the development of methods to calculate the Green's function using quantum devices with limited hardware resources has been actively pursued~\cite{endo2020calculation, rungger2019dynamical, chen2021variational, jamet2021krylov}.

In quantum embedding simulations, performing calculations in imaginary time rather than in real time is favorable for quantum computation with quantum devices in terms of the required quantum hardware resources.
This is because the imaginary-time formalism allows us to discretize the environment with fewer auxiliary bath sites~\cite{Bravyi:2017cc,Nusspickel2020,sparsebathfitting2021} and hence qubits.
Although recently proposed algorithms has been used to compute the imaginary-frequency Green's function~\cite{chen2021variational,PRXQuantum.2.010317, jamet2021krylov},
it requires the evaluation of the expectation value of the square of the Hamiltonian or an effective Hamiltonian,
which may be expensive for a large impurity model.
Thus, efficient methods for computing the imaginary-time Green's function using quantum devices with limited hardware resources still remain to be explored.

In this study, we propose a quantum--classical hybrid algorithm to compute the imaginary-time Green's function on quantum devices with limited hardware resource devices.
We apply a variational quantum simulation (VQS)~\cite{mcardle2019variational}, which is a variational algorithm for time evolution,
to the computation of the imaginary-time Green's function. 
Our algorithm efficiently computes the imaginary-time Green's function by adopting a non-uniform mesh focusing on the shape of the imaginary-time Green's function.
Using a quantum circuit simulator, we verified our method for typical impurity models, such as
a dimer model and a four-site impurity model obtained by DMFT calculations of the single-band Hubbard model.

The remainder of the paper is structured as follows. 
In Sec.~\ref{sec:green functions}, we review the finite-temperature and zero-temperature formalism of Green's functions. 
In Sec.~\ref{sec:vqa for green func}, we propose a variational quantum algorithm for computing the imaginary-time Green's function.
In Sec.~\ref{sec:results},
we numerically demonstrate the proposed algorithm for
typical impurity models using a quantum circuit simulator.
Finally, we discuss the scalability and numerical stability of our algorithm, and compare it with other methods.

\section{Green's functions}\label{sec:green functions}

\subsection{Finite-temperature formalism}\label{sec:finite-temperature}
We consider a fermionic system in the grand-canonical ensemble 
and denote its Hamiltonian as $\mathcal{ H}$.

\begin{align}
   \label{eq:hamiltonian}
   \mathcal{ H} &= \sum_{ij}^N t_{ij} \hat{c}^\dagger_i \hat{c}_j + \frac{1}{4} \sum_{ijkl} U_{ikjl}\hat{c}_i^\dagger \hat{c}_j^\dagger \hat{c}_l \hat{c}_k - \mu \sum_i \hat{c}^\dagger_i \hat{c}_i,
\end{align}
where $c_i$/$c^\dagger_i$ denote the creation and annihilation operators of the spin orbital $i$, and $N$ is the number of spin orbitals.
$t_{ij}$, $U_{ikjl}$, and $\mu$ denote the hopping matrix, Coulomb tensor, and chemical potential, respectively.
The retarded Green's function is defined as
\begin{align}
	G^\mathrm{R}_{ab}(t) &= -\ii \theta(t)\expval{\hat c_a(t)\hat c^\dagger_b(0) + \hat c^\dagger_b(0) \hat c_a(t)},
\end{align}
where $\hat c_a(t) = e^{\ii \mathcal{ H} t} \hat c_a e^{-\ii \mathcal{ H} t}$ and $\hat c^\dagger_b(t) = e^{\ii \mathcal{ H} t}\hat c^\dagger_b e^{-\ii \mathcal{ H} t}$ denote the annihilation and creation operators for spin orbitals $a$ and $b$ in the Heisenberg representation, respectively.
$\theta(t)$ is a step function.
Throughout this paper, we take $\hbar=k_\mathrm{B}=1$.
The thermal expectation $\expval{\cdots}$ is evaluated in the ground-canonical ensemble.
The retarded Green's function can be transformed into the (real) frequency space as
\begin{align}
G^\mathrm{R}(\omega) &= \int_{-\infty}^\infty \dd t~e^{\ii\omega t}G^\mathrm{R}(t),
\end{align}
where $\omega$ is a real frequency.

On the contrary, the imaginary-time Green's function is defined as
\begin{align}
    G_{ab}(\tau) &= -\theta(\tau)\expval{\hat c_a(\tau)\hat c^\dagger_b(0)}+
    \theta(-\tau)\expval{\hat c^\dagger_b(0)\hat c_a(\tau)},\label{eq:gtau}
\end{align}
where $\hat c_a(\tau) = e^{\tau \mathcal{ H}} \hat c_a e^{-\tau \mathcal{ H}}$.
Notably, the imaginary-time Green's function is anti-periodic, as $G_{ab}(\tau+\beta) = -G_{ab}(\tau)$.
The Fourier transform of the imaginary-time Green's function (Matsubara Green's function) is given by

\begin{align}
     G_{ab}(i\omega) &= \int_0^\beta \dd \tau e^{i\omega \tau}G_{ab}(\tau),
\end{align}
where $\omega = (2n+1)\pi/\beta$, where $n\in \mathbb{N}$ and $\beta = 1/T$.

The imaginary-frequency Green's function $G(i\omega)$ can be analytically continued from the imaginary axis to the full complex plane as $G_{ab}(z)$.
The analytically continued $G_{ab}(z)$ has the spectrum representation
\begin{align}
    &G_{ab}(z)=
   \int_{-\infty}^\infty \dd \omega \frac{\rho_{ab}(\omega)}{z - \omega}
\end{align}
with
\begin{align}
    &\rho_{ab}(\omega) \equiv 
    \sum_{mn}(e^{-\beta E_n} + e^{-\beta E_m})\times \nonumber \\
     &\mel{n}{\hat c_a}{m}\mel{m}{\hat c^\dagger_b}{n}
     \delta(\omega - (E_m-E_n)),\label{eq:matsubara-lehmann}
\end{align}
where $z$ is a complex number and
$n$, $m$ runs over all eigenstates of the system ($E_{m/n}$ denotes an eigenvalue).
There are poles for a finite system or a branch cut for an infinite system on the real axis.
The retarded Green's function is $G_{ab}(z)$'s value  just above the real axis:
\begin{align}
   G^\mathrm{R}_{ab}(\omega) &= G_{ab}(\omega+\ii 0^+).
\end{align}

\subsection{Zero-temperature formalism}\label{sec:zero-temperature}
We now consider the limit of $T\rightarrow 0$, where the ensemble average can be restricted to the ground state(s) $\Psi_\mathrm{G}$.
At sufficiently low  temperatures,
for $0 < \tau < \beta/2$, Eq.~\eqref{eq:gtau} can be rewritten as
\begin{align}
    G_{ab}(\tau) &= -\expval{\hat c_a(\tau)\hat c^\dagger_b(0)} \nonumber \\
    &\underset{T\rightarrow 0}{=} -\mel{\Psi_\mathrm{G}}{\hat c_a e^{-(\mathcal{ H}-E_\mathrm{G})\tau} \hat c^\dagger_b}{\Psi_\mathrm{G}},
    \label{eq:gtau-zero-T}
\end{align}
where $E_{\mathrm{G}}= \mel{\Psi_{\mathrm{G}}}{\mathcal{ H}}{\Psi_{\mathrm{G}}}$.
If the ground-state manifold is degenerate, then Eq.~\eqref{eq:gtau-zero-T} must be averaged over all degenerate ground states.
In general, $|G_{ab}(\tau)|$ decays exponentially for an insulating system, whereas its decay is algebraic for a metallic system.
We must increase $\beta$, which sets the upper limit of the time evolution such that $G_{ab}(\tau)$ is sufficiently small at the boundary.

Similarly, for $-\beta/2 < \tau < 0$, we obtain
\begin{align}
    G_{ab}(\tau) &\underset{T\rightarrow 0}{=}  \mel{\Psi_\mathrm{G}}{\hat c^\dagger_b e^{(\mathcal{ H}-E_{\mathrm{G}})\tau} \hat c_a}{\Psi_\mathrm{G}}.
    \label{eq:gtau-zero-T-negative-tau}
\end{align}
 Equations~\eqref{eq:gtau-zero-T} and \eqref{eq:gtau-zero-T-negative-tau} can be represented in a unified form as follows:
\begin{align}
    G_{ab}(\tau) &\underset{T\rightarrow 0}{=} \mp \mel{\Psi_\mathrm{G}}{ \hat A_\pm  e^{\mp (\mathcal{ H}-E_\mathrm{G})\tau}  \hat B_{\pm}}{\Psi_\mathrm{G}},\label{eq:gtau-zero-T-both}
\end{align}
where 
${A}_+ = \hat c_a$ and ${B}_+ = \hat c^\dagger_b$ for $\tau>0$, and
${A}_- = \hat c^\dagger_b$ and ${B}_- = \hat c_a$ for $\tau<0$.
The signs $\mp$ are for $\tau>0$ and  $\tau<0$, respectively.
Once $G_{ab}(\tau)$ is evaluated on a sufficiently fine mesh in $[-\beta/2,\beta/2]$ for a sufficiently large \textit{fictitious} inverse temperature $\beta$,
we can transform the data to the imaginary-frequency space, where the DMFT calculations are usually implemented, by numerically evaluating the integral
\begin{align}
    G_{ab}(i\omega) &= \int_{-\beta/2}^{\beta/2} \dd \tau e^{i\omega \tau}G_{ab}(\tau).\label{eq:matsubara}
\end{align}

\section{Variational quantum algorithms for computing Green's functions}\label{sec:vqa for green func}

\subsection{Overview}
\begin{figure*}
    \centering
    \vspace{-35mm} 
    \includegraphics[width=0.95\linewidth]{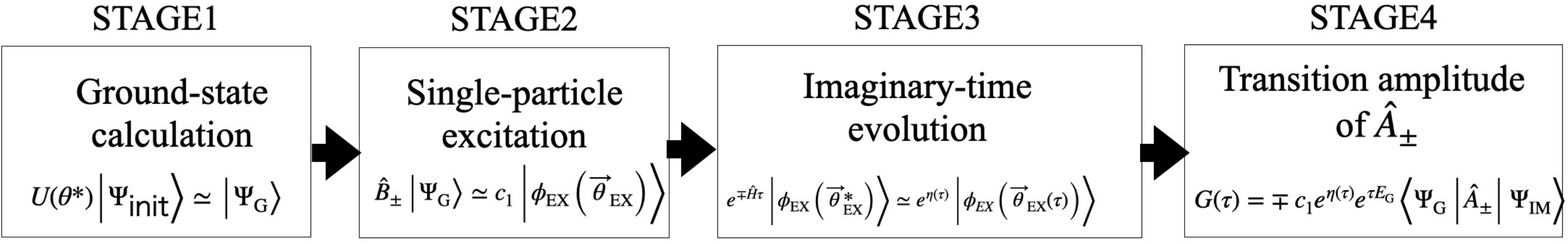}
    \vspace{-3mm}
    \vspace{-40mm} 
    \caption{
    Overview of the methods for computing the imaginary-time Green's function. 
    } 
    \label{fig:gf_stage}
\end{figure*}

We propose a variational algorithm for computing the imaginary-time Green's function.
First, we choose $\beta$ and introduce a fine mesh in $[-\beta/2, \beta/2]$.
Subsequently, we compute the Green's function for the positive and negative sides of $\tau$ with Eq.~\eqref{eq:gtau-zero-T-both} using the following procedure:

\begin{description}
    \item [Stage 1] The ground state is computed by a variational quantum algorithm.
    \item[Stage 2] The operator $\hat B_\pm$ is applied to the ground state.
    \item [Stage 3] Imaginary-time evolution is performed by VQS.
    \item [Stage 4] The Green's function is evaluated by computing the transition amplitude for $\hat A_\pm$.
\end{description}
Figure~\ref{fig:gf_stage} illustrates the entire  procedure.
In the following subsections, 
we explain the stepwise details of the procedure.
This method can be applied to general two-point correlation functions in the form of $\mel{\Psi_\mathrm{G}}{\hat A({\tau})  \hat B(0)}{\Psi_\mathrm{G}}$, where $\hat A$/$\hat B$ is an equal-time bosonic or fermionic operator. 

\subsection{Mapping to qubits}
To calculate a fermionic system on a quantum computer, it is necessary to convert the fermionic operators from a fermionic second quantized representation to a qubit representation in advance.
Typical methods include the Jordan--Wigner transformation~\cite{1928ZPhy...47..631J} and Braviy--Kitaev transformation~\cite{bravyi2002fermionic, seeley2012bravyi}.

As an example, we consider the Hamiltonian in the form of Eq.~\eqref{eq:hamiltonian}.
The Hamiltonian can be transformed to the qubit representation as
\begin{align}
    \label{eq:op_qubit_rep}
    \mathcal{ H} &\rightarrow \sum_{k} h_{k} {S}_{k},
\end{align}
where ${S}_{k}\in\{X, Y, Z, I\}^{\otimes m}$ are the tensor products of Pauli operators with $m$ qubits, which are transformed from the terms in Eq.~\eqref{eq:hamiltonian}; $h_p$ are the coefficients.
In this study, we adopt the Jordan--Wigner transformation given by
\begin{align}
    \label{eq:jw_1}
    &\hat c_{j}^{\dagger} \rightarrow \frac{1}{2}\left(X_{j}-i Y_{j}\right) Z_{1} Z_{2} \cdots Z_{j-1}, \\
    \label{eq:jw_2}
    &\hat c_{j} \rightarrow \frac{1}{2}\left(X_{j}+i Y_{j}\right) Z_{1} Z_{2} \cdots Z_{j-1},
\end{align}

\subsection{Stage 1: Ground-state calculations}\label{sec:vqe}
We use the variational quantum eigensolver (VQE)~\cite{peruzzo2014variational} to compute the ground state. VQE is a variational quantum algorithm for determining the ground state and its energy of the Hamiltonian in Eq.~\eqref{eq:op_qubit_rep} using a quantum computer.

The flow of the algorithm is described as follows.
\begin{description}
    \item[Step 1] Prepare an initial state $\ket{\Psi_\text{init}}$ (usually an unentangled product state) on a quantum computer.
    \item[Step 2] Generate a variational quantum state $\ket{\Psi(\vec{\theta^{}}))}$ (``ansatz'') by applying a unitary operator  $U(\vec{\theta^{}})$ with parameters $\vec{\theta^{}}$ to the initial state $\ket{\Psi_\text{init}}$.
    \item[Step 3] Measure $\expval{{S}_{k}}$ in the Hamiltonian of Eq.~\eqref{eq:op_qubit_rep} using a quantum computer.
    \item[Step 4] Calculate $\expval{\mathcal{ H}}$ on a classical computer.
    \item[Step 5] Update the parameters on the classical computer to reduce $\expval{\mathcal{ H}}$.
\end{description}
By repeating steps 2--5, we obtain a set of converged parameters $\vec{\theta}^{*}$.
If the representation capability of the ansatz is sufficiently high and an appropriate initial guess  is used, the optimized variational quantum state $\ket{\Psi(\vec{\theta}^*))}$ should approximate the ground state $\ket{\Psi_\mathrm{G}}$ accurately.

Throughout this paper, we assume that the quantum circuit conserves the number of electrons.
The number of electrons in the Hilbert space to be searched can be fixed by the number of electrons in the initial state and a number-conserving circuit.

In step 5, if we update the parameters using the gradient method, evaluating the derivative of $\expval{\mathcal{ H}}$ is necessary.
This can be done on a quantum computer either by using numerical finite differentiation or parameter-shift rules~\cite{Mitarai.2018.Fujii}.

\subsection{Stage 2: Single-particle excitation}

In stage 2, we compute a variational quantum state for the single-particle excited state ${\hat B}_\pm \ket{\Psi_\mathrm{G}}$.
Because the operator is not unitary, we represent the resultant state as
\begin{align}
    \label{eq:ex_state}
   {\hat B}_\pm \ket{\Psi_\mathrm{G}} &\simeq c_1 \ket{\phi_\mathrm{EX}(\vec{\theta}_\mathrm{EX})},
\end{align}
where $c_1$ is a coefficient and the parametrized quantum state $\ket{\phi_\mathrm{EX}(\vec{\theta}_\mathrm{EX})}$ is defined by
\begin{align}
\ket{\phi_\mathrm{EX}(\vec{\theta}_\mathrm{EX})} &= U(\vec{\theta}_\mathrm{EX}) \ket{\phi_\mathrm{EX}},
\end{align}
where the initial state $\ket{\phi_\mathrm{EX}}$ must have $N\pm 1$ electrons because the operator changes the number of electrons by $\pm 1$.

We compute the variational parameters $\vec{\theta}_\mathrm{EX}$ and coefficient $c_1$ as follows:
\begin{description}
    \item[Step 1] 
    Transform the creation operator 
    ${\hat B}_\pm$
    to the qubit representation, and prepare an initial guess for $\vec{\theta}_\mathrm{EX}$.
    \item[Step 2]
    Prepare a variational quantum state 
    $\phi_\mathrm{EX}(\vec{\theta}_\mathrm{EX})$ 
    and measure the cost function 
    $C=\left|\mel{\phi_\mathrm{EX}(\vec{\theta}_\mathrm{EX})}{\hat B_\pm}{\Psi_\mathrm{G }}\right|^{2}$ 
    on a quantum computer. 
    We repeatedly optimize the parameters to reduce the cost function until the parameters converge.　
    \item[Step 3]
    For the converged parameters $\vec{\theta}_\mathrm{EX}^*$,
    measure $c_1$=$\mel{\phi_\mathrm{EX}(\vec{\theta}^{*}_\mathrm{EX})}{\hat B_\pm}{\Psi_\mathrm{G}}$ on a quantum computer.
\end{description}

In steps 2 and 3, we evaluate the cost function and $c_1$ on a quantum circuit.
We explain the method of evaluating $c_1$ in step 3 because the cost function in step 2 can be directly obtained by squaring the absolute value of $c_1$
~\footnote{In \cite{ibe2022calculating}, it was proposed to reduce the evaluation of the transition amplitudes of the form  $|\mel{0} {U(\vec{\theta_a})PU(\vec{\theta}_b)}{0}|^{2}$ to the sum of measurements of the overlap of two states when $P$ is Hermitian. If this technique can be extended to the case where $P$ is not only a Hermitian operator but also a general operator such as the fermionic creation and annihilation operator, the same technique may be used herein.}.
\begin{figure*}
    \centering
    \vspace{-20mm} 
    \includegraphics[width=0.85\linewidth]{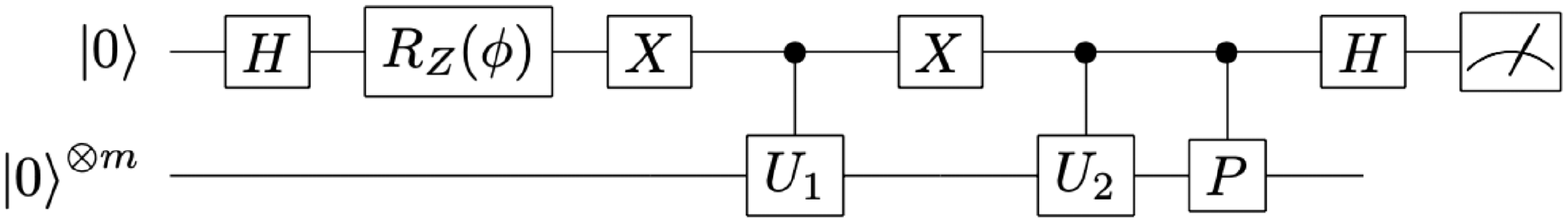}
    \vspace{-3mm}
    \vspace{-40mm} 
    \caption{Quantum circuit for calculating Eq.~(\ref{general_transition_amplitude}).
    This quantum circuit uses $m$ qubits (bottom line) and one ancilla qubit (top line). The transition amplitude can be calculated by combining the $Z$ measurement results for the ancilla qubit.
    } 
    \label{fig:transition_amplitude}
\end{figure*}

First, we decompose ${B}_\pm$ as the sum of its Hermitian part and its anti-Hermitian part using the Jordan--Wigner transformation, as in Eq.~\eqref{eq:jw_1}, and Eq.~\eqref{eq:jw_2}.
We evaluate the transition amplitude on a quantum computer by measuring the Hermitian and anti-Hermitian parts of the following form:
\begin{align}
    \label{general_transition_amplitude}
    \mel{0} {U_{1}^{\dagger}{P}_{}U_{2}}{0},
\end{align}where $P$ are Pauli operators with $m$ qubits, and $U_1$ and $U_{2}$ are unitary operators with $m$ qubits. Equation~\eqref{general_transition_amplitude} can be measured using the quantum circuit in Fig.~\ref{fig:transition_amplitude}~\cite{endo2020calculation,chen2021variational},
which requires one ancilla qubit.

Let $p_{0}$/$p_{1}$ be the probability of measuring 0/1 in the ancilla qubit.
The real and imaginary parts of the transition amplitude can be measured separately by setting $\phi=0$ and $\pi/2$ in the $R_z$ gate, respectively, as
\begin{align}
p_{0} - p_{1} &=
\begin{cases}
\Re\mel{0} {U_{1}^{\dagger}(\vec{\theta_1}) P U_{2}(\vec{\theta}_2)}{0}  & \phi=0,\\
-\Im\mel{0}{U_{1}^{\dagger}(\vec{\theta_1}) P U_{2}(\vec{\theta}_2)}{0} & \phi=\pi/2.
\end{cases}
\end{align}
As this method is based on a single ancilla qubit,
we need complex quantum circuits owing to  the use of the control unitary operators .

If we use a derivative-based optimization algorithm in step 2, 
we additionally need to measure the partial derivatives of $C$ with respect to $\vec{\theta}_{EX}$. 
The analytical form of the derivative of this cost function can be calculated using the quantum circuit of the same form to evaluate the transition amplitude.

In the following, we use the same quantum circuit to evaluate the quantities in the same form as Eq.~(\ref{general_transition_amplitude}) in stages 3 and 4.

\subsection{Stage 3: Imaginary-time evolution}\label{sec:vqs}
In stage 3, we calculate the imaginary-time evolution of $\ket{\phi_\mathrm{EX}(\vec{\theta}_\mathrm{EX}^*)}$ obtained in stage 2.
As the imaginary-time evolution is not unitary, we represent the imaginary-time evolved state~\footnote{In general, one can use a more general quantum circuit for imaginary-time evolved states for $\tau>0$ than that used to fit the single-particle excited state in Stage 2. For simplicity, in this study, we used the same quantum circuit.} as
\begin{align}
    \label{eq:ex_state}
    e^{- H \tau}  \ket{\phi_\mathrm{EX}(\vec{\theta}_\mathrm{EX}^*)} &\simeq e^{\eta(\tau)} \ket{\phi_{EX}(\vec \theta_{\mathrm{EX}} (\tau))},
\end{align}
where $\eta(\tau)$ is a $\tau$-dependent real number.
A similar parametrization for nonunitary time evolution was proposed for the application of VQS to financial systems~\cite{PhysRevA.103.052425}.
Also, a different normalization factor calculation method was proposed in the context of calculation of the Gibbs partition function, which was the first paper to calculate the normalization factor~\cite{matsumoto2022calculation}.

We perform the imaginary-time evolution on a (generally non-uniform) mesh, $\qty{\tau_1, \tau_2, \cdots, \tau_N}$ ($\tau_1=0$ and $\tau_N=\beta/2$).
This can be achieved by determining the variational parameters $\vec \theta_\mathrm{EX}(\tau_t)$ and $\eta(\tau_t)$ on the mesh points sequentially from $\tau=0$ to $\tau=\beta/2$.

\subsubsection{Time-dependent variational principle}\label{sec:tdvp}
We first review the time-dependent variational principle (TDVP), on which VQS is based.
The time-dependent Schr\"{o}dinger equation reads
\begin{align}
    \label{eq:schrodinger_eq}
    \dv{\tau}\ket{\Psi(\tau)} &= -\mathcal{ H} \ket{\Psi(\tau)},
\end{align}
where $\tau$ is an imaginary time. 
The imaginary-time evolution of the normalized ket obeys
\begin{align}
    \dv{\tau}\ket{\tilde \Psi(\tau)} &= -(\mathcal{ H} - E_\tau) \ket{\tilde \Psi(\tau)},\label{eq:vqs-eq}
\end{align}
where $\ket{\tilde \Psi(\tau)} \equiv \ket{\Psi(\tau)}/\sqrt{\bra{\Psi(\tau)}\ket{\Psi(\tau)}}$ and $E_\tau \equiv \expval{\mathcal{ H}}{\tilde \Psi(\tau)}$.

We now parametrize these two kets as
\begin{align}
    \label{eq:imag_state1}
    \ket{\tilde{\Psi}(\tau)} &= \ket{\phi(\vec \theta (\tau))},\\
    \label{eq:imag_state2}
    \ket{\Psi(\tau)} &= e^{\eta(\tau)} \ket{\phi(\vec \theta (\tau))},
\end{align}
where $\bra{\phi}\ket{\phi} = 1$, the vector $\vec \theta (\tau)$ denotes the $\tau$-dependent real variational parameters, and $\eta(\tau)$ is a real parameter for the norm.

In the TDVP, the time evolution of Eq.~\eqref{eq:vqs-eq} is mapped to the time evolution of $\vec{\theta}(\tau)$.
In McLachlan's variational principle, we minimize the distance between the exact evolution and the evolution of the parametrized state under infinitesimal variation of the imaginary time $\delta \tau$ as
\begin{align}
    \mathrm{min}~\delta \norm{ \qty(\dv{\tau} + \mathcal{ H} - E_\tau) \ket{\phi(\vec{\theta}(\tau)}},\label{eq:min-vqs}
\end{align}
where $\norm{\cdots}$ denotes the Frobenius norm.
There are several possible ways to solve this equation.
We will explain them later .

Once the $\tau$ dependence of $\vec \theta$ is determined,
we can compute $\eta(\tau)$ by solving the differential equation 
\begin{align}
  \dv{\eta(\tau)}{\tau} &= - E_\tau\label{eq:d_eta}
\end{align}
with an appropriate initial condition at $\tau=0$.
Equation~\eqref{eq:d_eta} can be derived by
substituting Eq.~(\ref{eq:imag_state2}) into Eq.~(\ref{eq:schrodinger_eq}).

\subsubsection{VQS and Direct VQS}\label{sec:VQS}
\subsubsection*{VQS}
We explain how to perform the imaginary-time evolution of a quantum state on a discrete mesh in $\tau$ in VQS. 
Here, we define $\ket{\phi(\vec{\theta}(\tau)}$ as $U(\vec{\theta}) \ket{\phi}$ on the quantum circuit, where $U(\vec{\theta(\tau)})$ is a unitary operator with $\tau$-dependent real parameters and $\ket{\phi}$ is an initial state. 
One can explicitly write the equation for determining the time derivative of the variational parameters at $\tau$:
\begin{align}
\label{eq:derivative_parameters}
    \sum_{j=1}^{N_\mathrm{P}} M_{ij} \dot{\theta}_j(\tau) &= C_i,
\end{align}
where
\begin{align}
\label{eq:Aij}
    M_{ij} 
    &\equiv
    \mathcal{R} \pdv {\bra{\phi}U^{\dagger}(\vec \theta(\tau))}{\theta_i}
    \pdv{ U(\vec \theta(\tau))\ket{\phi}}{\theta_j}. 
    \\
    \label{eq:Ci}
    C_i 
    &\equiv
    - \mathcal{R} \bra{\phi}U^{\dagger}(\vec \theta(\tau)) \mathcal{ H}
    \qty
    (
      \pdv{U(\vec \theta(\tau))\ket{\phi}}{\theta_i}
     )
\end{align}
and $N_\mathrm{P}$ denotes the number of variational parameters. 

Equations.~(\ref{eq:Aij}) and (\ref{eq:Ci}) involve quantum circuits differentiated with respect to the variational parameters.
As discussed in detail in Ref.~\onlinecite{matsumoto2022calculation},
the matrix elements of $M$ and $C$ can be efficiently computed on a quantum computer using one ancilla qubit, for example, by 
differentiating the variational quantum circuits explicitly
(see Appendix~\ref{sec:measurement circuit of vqs} for more details).

The linear system in Eq.~\eqref{eq:derivative_parameters} can be solved efficiently on a classical computer~\footnote{In the present study, we use the subroutine gelsy based on QR decomposition in LAPACK to solve the linear system. In the case of redundancy in the parameterization of a quantum circuit, we remove small singular values.}.
Subsequently, we evolve the quantum state from $\tau$ to $\tau+\Delta \tau$ ($\Delta \tau>0$) by updating the variational parameters as
\begin{align}
    \label{eq:next_theta_vqs}
    \theta_i(\tau+\Delta\tau) &\simeq \theta_i(\tau) + \Delta \tau \dot{\theta}_i(\tau).
\end{align}

\subsubsection*{Direct VQS}
Here, we propose an alternative way to perform the imaginary-time evolution, which is based on the direct minimization of Eq.~\eqref{eq:min-vqs} for a finite time step, $\Delta \tau$.
We name this approach ``direct VQS''.
The optimization problem in Eq.~\eqref{eq:min-vqs} is equivalent to the following optimization problem:
\begin{align}
  \label{eq:direct}
   &\vec{\theta}(\tau+\Delta \tau)\nonumber\\
  & \simeq \underset{\vec{\theta}}{\mathrm{argmin}} \norm{\ket{\phi(\vec{\theta})} - \ket{\tilde{\Psi}(\tau)} + \Delta\tau (\mathcal{ H}-E_\tau) \ket{\tilde{\Psi}(\tau)}}\\
  \label{eq:direct_}
 &=\underset{\vec{\theta}}{\mathrm{argmin}} \Re[\Delta\tau\mel{\phi(\vec{\theta})}{\mathcal{ H}}{\tilde{\Psi}(\tau)}\nonumber\\
 &\hspace{1em}-(\Delta\tau E_\tau + 1) \braket{\phi(\vec{\theta})}{\Psi(\tau)}],
\end{align}
where $\ket{\tilde{\Psi}(\tau)} = \ket{\phi(\vec{\theta}(\tau)}$.
The terms that do not depend on $\vec{\theta}(\tau + \Delta \tau)$ are excluded from the cost function.
Equation~\eqref{eq:direct} becomes exact in the limit $\Delta \tau \rightarrow 0$.
This optimization problem can be efficiently solved using $\vec{\theta(\tau)}$ as an initial guess.
We can evaluate the quantities in Eq.~\eqref{eq:direct_} by using a quantum circuit as Eq.~\eqref{general_transition_amplitude}, 
after decomposing the Hamiltonian into a sum of terms in the form of Eq.~\eqref{eq:op_qubit_rep}.

\subsubsection{Computational complexity of VQS and direct VQS}\label{sec:computational complexity}
The computational complexity of VQS and direct VQS depends on the Hamiltonian and the ansatz.
We discuss the computational complexity of VQS and direct VQS for a Hamiltonian with a general two-body interaction and a quantum impurity model.
In this discussion, we assume the unitary coupled cluster ansatz with generalized singles and doubles (UCCGSD)~\cite{nooijen2000can, lee2018generalized} (see Appendix~\ref{sec:details for vqa} for more details).
For this ansatz, the number of variational parameters $N_{\mathcal{P}}$ and the gate depth scale as $O(n_{\mathrm{so}}^4)$  [$n_{\mathrm{so}}$ is the total number of spin orbitals]. 
After Jordan--Wigner transformation, the gate depth scales as $O(n_{\mathrm{so}}^5)$.
However, the gate depth reduces to $O(n_{\mathrm{so}}^4)$ if we can maximally parallelize the terms in the unitary coupled cluster operator on a near-term quantum computer \cite{o2019generalized}.

\subsubsection*{A Hamiltonian with a general two-body interaction}

In the case of a Hamiltonian with a general two-body interaction, the number of terms $N_{\mathrm{h}}$ in the Hamiltonian scales as $O(n_{\mathrm{so}}^4)$.

\textit{[VQS]} 
The bottleneck of imaginary-time evolution using VQS is the evaluation of the vector $C$ and the matrix $M$ on a quantum computer.
The computational complexity for measuring the elements of $C$ scales approximately as $O(N_{\mathrm{depth}}N_\mathrm{P}N_{\mathrm{h}})$, where $N_{\mathrm{depth}}$ is the depth of a quantum circuit.
In particular, 
for the UCCGSD ansatz, $O(N_{\mathrm{depth}}N_\mathrm{P}N_{\mathrm{h}}) \varpropto O(n_{\mathrm{so}}^{12})~[\because N_{\mathrm{depth}}=O(n_{\mathrm{so}}^{4})]$.
On the other hand, approximately, the computational complexity for measuring the elements of $M$ scales as $O(N_{\mathrm{depth}}N_\mathrm{P}^2) \varpropto O(n_{\mathrm{so}}^{12})$.
This is because 
we need to evaluate the quantum circuit(s) comprising $O(N_\mathrm{P})$ parameters to evaluate each matrix element~\cite{matsumoto2022calculation}.

\textit{[Direct VQS]} One does not have to evaluate the elements of the large matrix and vector, in contrast to the original VQS.
The computational complexity scales as $O(N_{\mathrm{depth}}N_\mathrm{P}N_\mathrm{h} N_\mathrm{iter}) \varpropto O(n_{\mathrm{so}}^{12} N_\mathrm{iter})$, where $N_\mathrm{iter}$ is the number of iterations required for optimization.
If $N_\mathrm{iter}$ does not depend strongly on $N_\mathrm{P}$, the direct VQS is as scalable as original VQS in terms of computational complexity.
One has to evaluate the first and second terms of the last line of Eq.~\eqref{eq:direct_} and subtract the second term from the first one.
This may be unstable under the influence of noise.

For comparing the computational complexity of VQE and that of direct VQS, 
both are roughly equivalent in terms of computation complexity 
because of the VQE's computational complexity $O(N_{\mathrm{depth}}N_\mathrm{P}N_\mathrm{h} N_\mathrm{iter}) \varpropto O(n_{\mathrm{so}}^{12} N_\mathrm{iter})$ with the UCCGSD ansatz.
A more efficient and compact ansatz with a better scaling is desired.

\subsubsection*{Quantum impurity models}
For a quantum impurity model with a \textit{starlike geometry} (see Fig.~\ref{fig:dmft}), $N_{\mathrm{h}}$ scales as $O(N_{\mathrm{bath}} + N_{\mathrm{bath}}N _{\mathrm{imp}}+N_{\mathrm{imp}}^4)$, where $N_{\mathrm{imp}}$ and $N_{\mathrm{bath}}$ are the number of impurity orbitals and the number of bath orbitals, respectively.

For discretizing a continuous bath, $N_{\mathrm{bath}} \varpropto N_{\mathrm{imp}}$ are known to suffice~\cite{Bravyi:2017cc, sparsebathfitting2021}.
Note that $N_{\mathrm{imp}} \ll N_{\mathrm{bath}}$ holds for quantum impurity models describing real materials.
For example, $N_{\mathrm{bath}}$ and $N_{\mathrm{imp}}$ required for a clustered DMFT calculation of the iron-based superconductor LaFeAsO were estimated to be $N_{\mathrm{imp}}$ = 40, $N_{\mathrm{bath}}$ = 332~\cite{sparsebathfitting2021}.
 For such a large impurity model, $N_{\mathrm{h}}$ scales as $O(N_{\mathrm{imp}}^4)$ with $N_\mathrm{{imp}}^4 \ll n_\mathrm{so}^4$.

\subsection{Stage 4: Computing the transition amplitude for $\hat A_\pm$}\label{sec:transition_amplitude_stage4}
In Stage 4, we compute the imaginary-time Green's function 
\begin{align}
     \label{eq:final_form_of_gr}
     G(\tau) = - c_{1} e^{\eta(\tau)} e^{\tau E_\mathrm{G}} \mel{\Psi_{\mathrm{G}}}{\hat A_\pm}{\Psi_{\mathrm{IM}}},
\end{align}
where $E_\mathrm{G}$ is the (approximate) ground-state energy obtained by VQE.
We measure the quantity $\mel{\Psi_{\mathrm{G}}}{\hat A_\pm}{\Psi_{\mathrm{IM}}}$ by using the same circuit 
as Eq.~\eqref{general_transition_amplitude}.

\section{Numerical results}\label{sec:results}
As an application of the two proposed algorithms using VQS and direct VQS, we solve a dimer model and a four-site impurity model obtained by DMFT calculations for the Hubbard model using a quantum circuit simulator.

\subsection{Numerical details}
In this study, we used \texttt{Qulacs}~\cite{suzuki2021qulacs}, \texttt{pyed~}\cite{pyed}, \texttt{Openfermion}~\cite{mcclean2020openfermion}, and \texttt{irbasis}~\cite{chikano2019irbasis} to implement the proposed method.
\texttt{Qulacs} is used as a quantum circuit simulator. We use the \texttt{pyed} library, which is based on TRIQS (Toolbox for Research on Interacting Quantum Systems)~\cite{Parcollet:2015gq}, for computing the reference data of the Green's function.
\texttt{Openfermion} is used in the Jordan--Wigner transformation and to calculate the exact eigenvalues of models.
We generate a sparse sampling mesh for a sufficiently large $\beta=1000$ using \texttt{irbasis}
(see Appendix~\ref{sec:ft} for more details).
The sparse mesh covers $-\beta/2 < \tau < 0$ and $0 < \tau < \beta/2$,
when computing $G(\tau)$ for $\tau<0$ and $\tau>0$, respectively.
We perform DMFT calculations using \texttt{DCore}~\cite{shinaoka2021dcore} to generate the four-site impurity model.

To deal with the numerical instability of the computing imaginary-time Green's function, 
we solve Eq.~\eqref{eq:derivative_parameters} using a truncated singular value decomposition
as proposed in Ref.~\onlinecite{mcardle2019variational} and adopted additional tricks.
See Appendix~\ref{sec:safe_methods} for more details.

In the following results, we used the unitary coupled cluster ansatz with generalized singles and doubles (UCCGSD) as the quantum circuit.
We used the quasi-Newton method (the BFGS method) for optimizing the variational parameters.
The gradients of cost functions are computed by a finite difference method.
In Eq.~(\ref{eq:next_theta_vqs}), we set $\Delta \tau = 10^{-5}$.
In this study, we used random initial guesses as we observed that optimization was sometimes trapped in a metastable solution when starting from a zero initial guess.

In the following calculations, we used an MPI-parallelized program. We used a workstation equipped with an AMD EPYC 7702P 64-core processor. To solve the largest model where the total number of circuit parameters is 1568, VQS took 10 hours with 22 cores, whereas  direct VQS took 3 hours with 10 cores.

\subsection{Dimer model}\label{sec:dimer}
\begin{figure}[ht]
\centering
\vspace{-5mm} 
\includegraphics[width=0.8\linewidth]{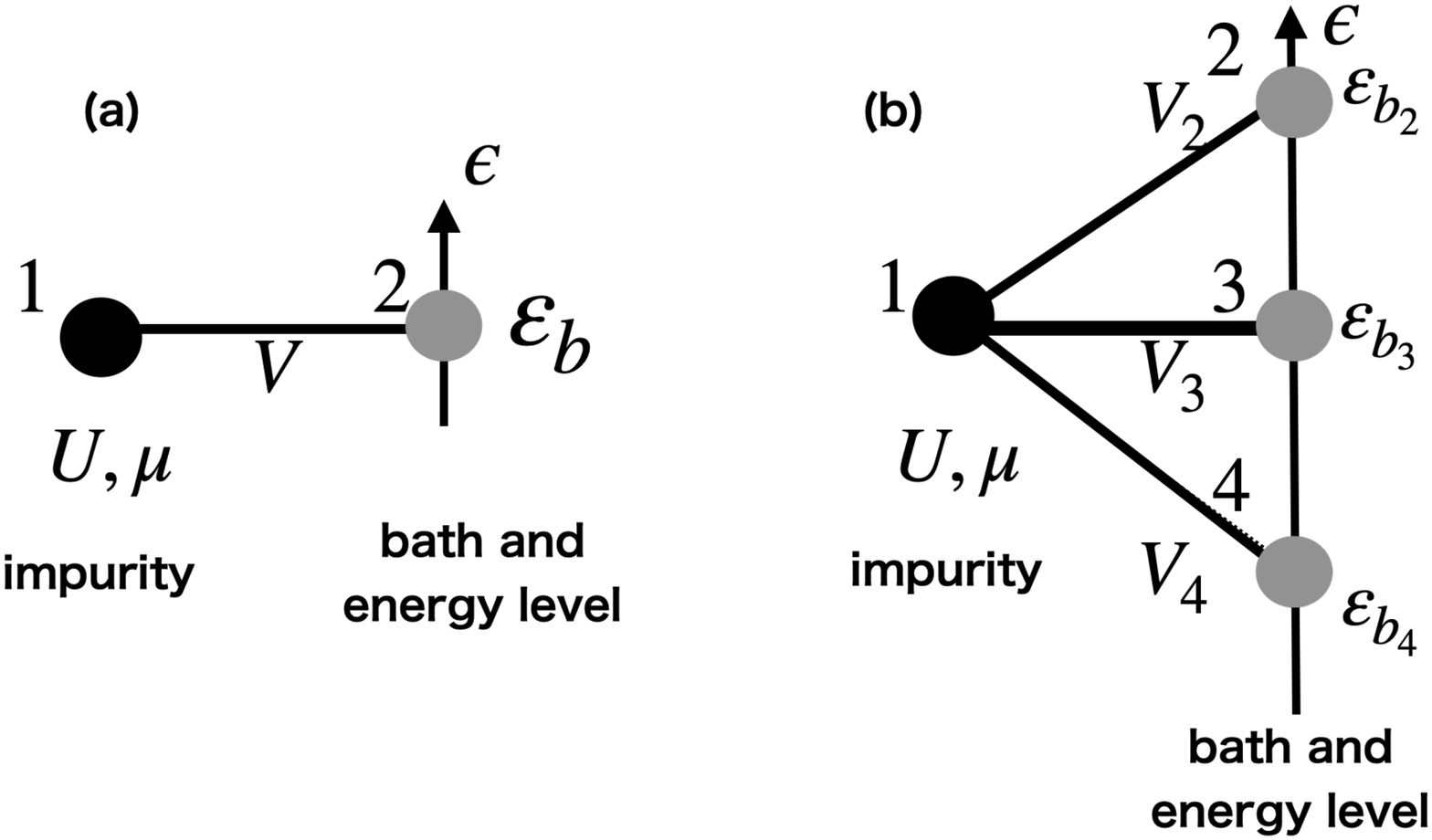}
\vspace{-5mm} 
\caption{(a) Dimer model and (b) four-site impurity model.}
\label{fig:dimer}
\end{figure}

We first demonstrate the algorithms with a dimer model. 
Its Hamiltonian reads
\begin{align}
     \mathcal{ H} &=U \hat{n}_{1 \uparrow} \hat{n}_{1 \downarrow}-\mu\sum_{\sigma=\uparrow, \downarrow}\hat{n}_{1 \sigma}
     \nonumber\\
     &-V\sum_{\sigma=\uparrow, \downarrow}
     \left(\hat{c}_{1 \sigma}^{\dagger} \hat{c}_{2 \sigma}+\hat{c}_{2 \sigma}^{\dagger} \hat{c}_{1 \sigma}\right)+\epsilon_{b}\sum_{\sigma=\uparrow, \downarrow}\ \hat{n}_{2 \sigma},
\end{align}
where $\hat{n}_{i \sigma} (\equiv\hat{c}_{i \sigma}^{\dagger} \hat{c}_{i \sigma})$ is the spin density operator at site $i$ and spin $\sigma$.
As illustrated in Fig.~ \ref{fig:dimer}, the dimer comprises one interacting ``impurity site'' with an onsite repulsion $U$ and one one-interacting ``bath site.''
This corresponds to the case where there is one impurity site and one bath site (Fig.~\ref{fig:dmft}).
We take $U=1$, $\mu = U/2$, $\epsilon_{b}= 1$ and $V = 1$.
This model is not particle--hole symmetric because $\epsilon_b \neq 0$.

The exact ground-state energies for the number of particles $n=1,2,3,4$ are $-1,-1.5,0.2,2$, respectively.
Thus, the global ground state has $n=2$.
In particular, there is a finite large energy gap between the ground and excited states.

\begin{figure}
    \includegraphics[width=0.75\linewidth]{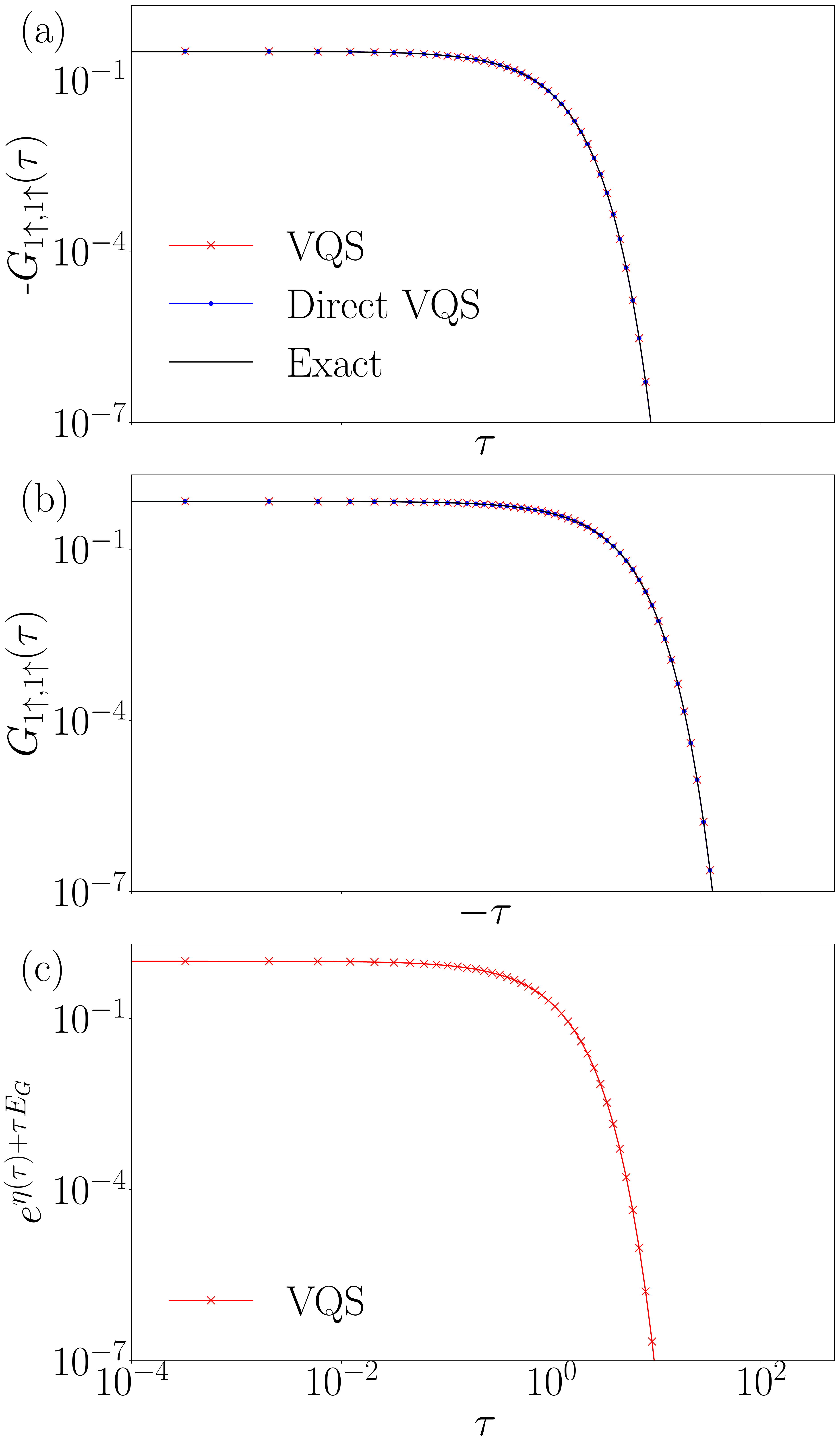}
    \caption{
    Computed $G_{1\uparrow,1\uparrow}(\tau)$ for the dimer model. 
    Panels (a) and (b) show the results for $\tau>0$ and $\tau<0$, respectively. Data smaller than $10^{-8}$ are not shown.
    Panel (c) shows the result of the exponential component of Green's function ($\tau>0$).
    }
    \label{fig:dimer_vqs_direct_GF_plots}
\end{figure}

Figures~\ref{fig:dimer_vqs_direct_GF_plots}(a) and \ref{fig:dimer_vqs_direct_GF_plots}(b) show 
$G_{1\uparrow,1\uparrow}(\tau)$
computed by the VQS and the direct VQS, respectively.
We used 70 and 70 sparse sampling points for $\tau>0$ and $\tau<0$, respectively.
Green's functions computed by VQS and direct VQS are in good agreement with the exact result. 
The exponential component $e^{\eta(\tau)+\tau E_\mathrm{G}}$ is plotted in Fig.~\ref{fig:dimer_vqs_direct_GF_plots}(c) for $\tau>0$.
On comparing Figs.~\ref{fig:dimer_vqs_direct_GF_plots}(a) and (b),
one can see that this term mainly determines the exponential decay of $G(\tau)$.
As mentioned earlier, there is a large gap in energy around the number of particles 2,  which results in an exponential decay of the imaginary-time Green's function. This indicates that the system is insulating.

\begin{figure}[h]
    \centering
    \includegraphics[width=0.80\linewidth]{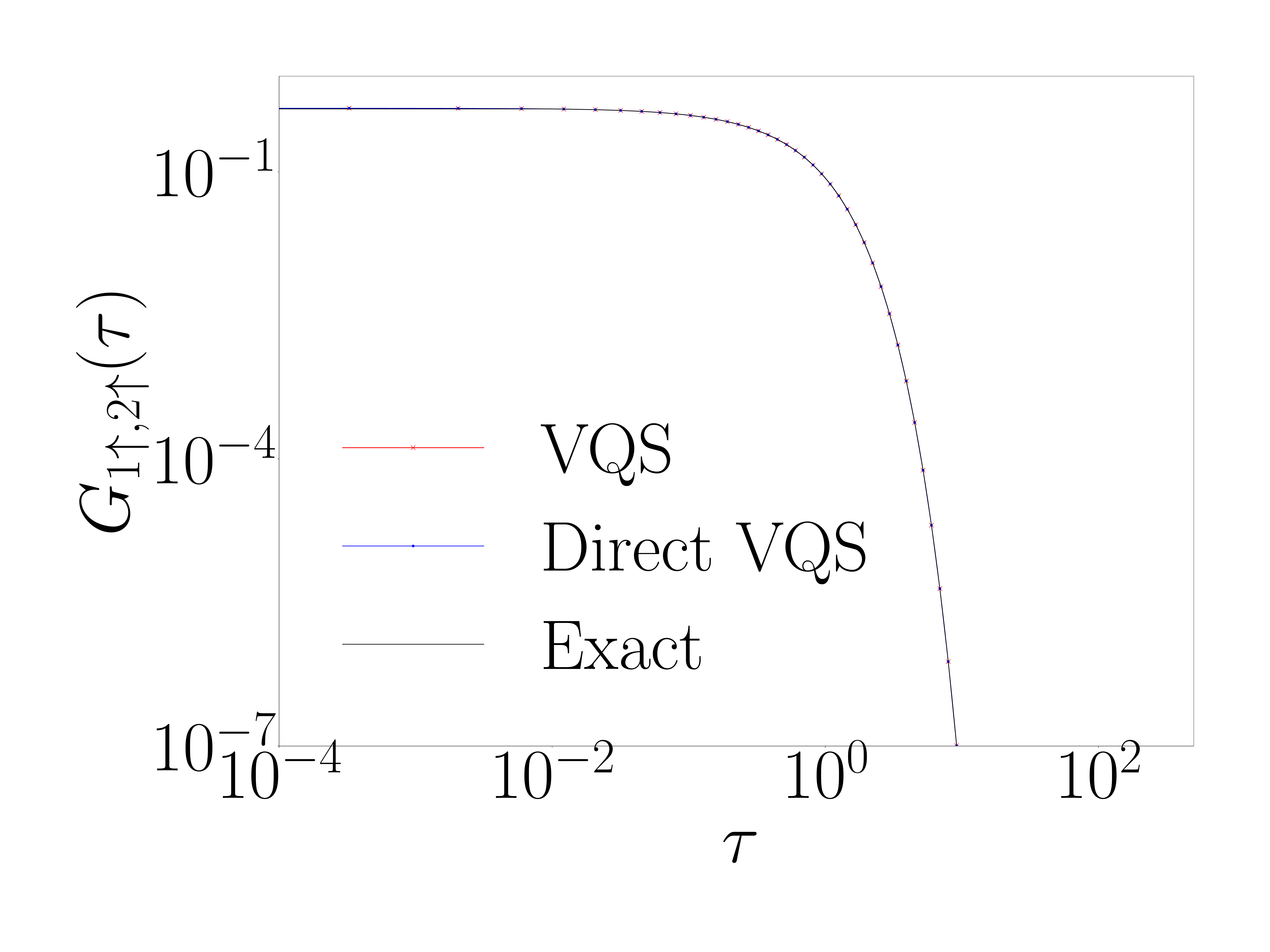}
    \caption[]{
    Off-diagonal component of Green's functions of the dimer model, $G_{1\uparrow,2\uparrow} (\tau)$, is computed for $\tau>0$.
    }
    \label{fig:dimer_offdiagonal_GF}
\end{figure}

Figure~\ref{fig:dimer_offdiagonal_GF} shows the off-diagonal component of the Green's function computed for $\tau>0$.
This was done using $\hat{A}_{+}=c_{1,\uparrow}$ and $\hat{B}_{+}=c_{2,\uparrow}^{\dagger}$.
The results clearly demonstrate that the off-diagonal component can be accurately measured using the proposed methods.

\begin{figure}[ht]
    \includegraphics[width=0.80\linewidth]{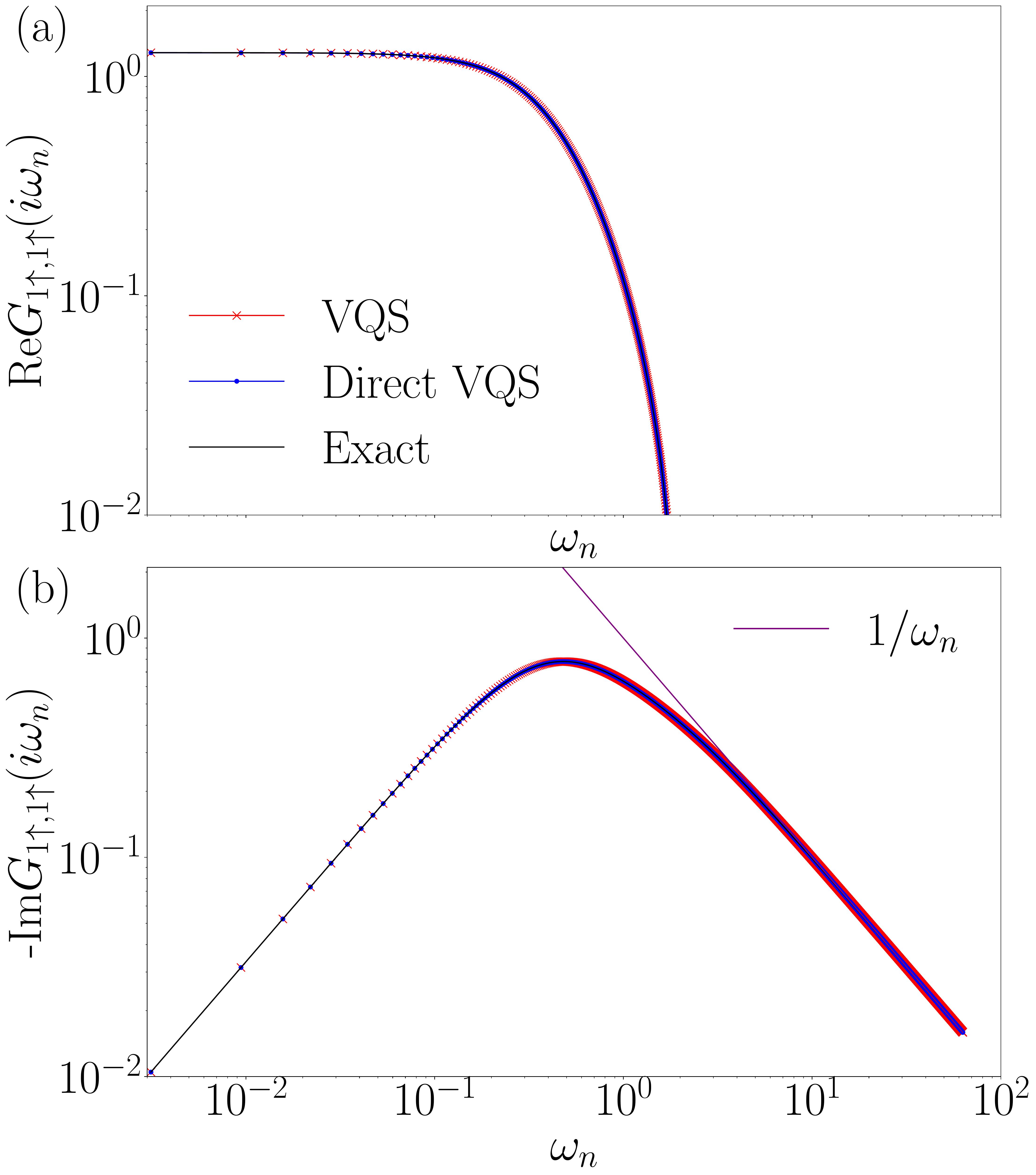}
    \caption[]{
    Matsubara Green's function computed for the dimer model.
    Panels (a) and (b) show the results for the real part and imaginary part, respectively.
    }
    \label{fig:dimer_GF_ft_plots}
\end{figure}

Figure~\ref{fig:dimer_GF_ft_plots} shows the Matsubara Green's function transformed from the τ domain.
The Matsubara Green's function computed by VQS and direct VQS agree with the exact result from low to high frequencies.
At high frequencies, the Green's function decays as $1/\mathrm{i}\omega_n$,
which is consistent with the fact that the Green's function has a discontinuity of 1 at $\tau=0$ owing to the non-commutativity of the creation and annihilation operators.

\subsection{Four-site model: Effective model in single-orbital DMFT}\label{sec:four-site}

Next, we consider the particle--hole symmetric four-site ``impurity'' model defined by the Hamiltonian
\begin{align}
     \mathcal{ H} &= U \hat{n}_{1 \uparrow} \hat{n}_{1 \downarrow}-\mu\sum_{\sigma=\uparrow, \downarrow}\hat{n}_{1 \sigma} \nonumber \\&
     -\sum_{k=2}^{4}\sum_{\sigma=\uparrow, \downarrow}
     V_{k}\left(\hat{c}_{1 \sigma}^{\dagger} \hat{c}_{k \sigma}+\hat{c}_{k \sigma}^{\dagger} \hat{c}_{1 \sigma}\right)+\epsilon_{k}
     \sum_{k=2}^{4}\sum_{\sigma=\uparrow, \downarrow}\hat{n}_{k \sigma},
\end{align}
where $\hat{n}_{i \sigma}=\hat{c}_{i \sigma}^{\dagger} \hat{c}_{i \sigma}( \sigma=\uparrow, \downarrow)$, and $k$ is an index for ``bath sites''. 
This model corresponds to the case where there is one impurity site and three bath sites (Fig.~\ref{fig:dmft}).
As shown in Fig.~\ref{fig:dimer}, the correlated impurity site is coupled to all three bath sites through the coupling terms $V_k = [0.0, -1.26264, 0.07702, -1.26264]$.
There is no direct coupling between different bath sites.
We take $\mu=U/2$, $\epsilon_{k} =[0.0, 1.11919, 0.0, -1.11919]$.
These parameters were determined by single-site DMFT calculations of the single-orbital Hubbard model on a square lattice with an onsite repulsion of $U=4$ at half filling and zero temperature.
The critical value of the Mott transition is $U\simeq 12$.
The exact ground-state energies of the model are $-3.16$, $-5.31$, $-5.49$, and $-5.51$ for the number of particles $n=1,2,3,4$.
Thus, there is only a small energy gap (0.02 between $n=4$ and $n=3,5$).

\begin{figure}[h]
    \includegraphics[width=1.0\linewidth]{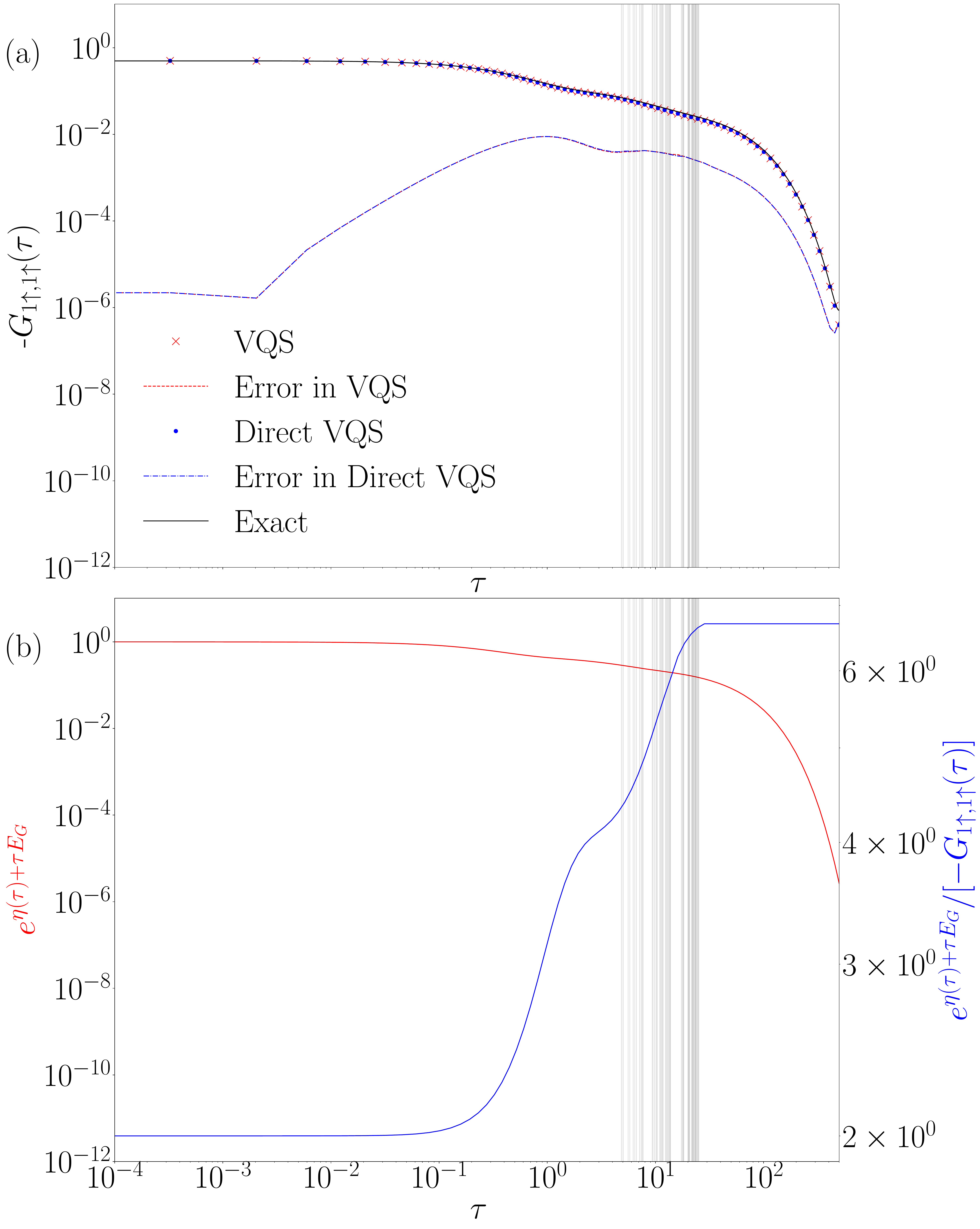}
    \caption{
    Computed $G_{1\uparrow,1\uparrow}(\tau)$ for the four-site impurity model.
    Panel (a) shows the result for $\tau>0$.
    The vertical thin lines denote the 
    additional imaginary-time points added to avoid numerical instability (see the main text and Appendix~\ref{sec:safe_methods}).
    Panel (b) shows different contributions to the Green's function ($\tau>0$).}
    \label{fig:4site_vqs_direct_GF_plots}
\end{figure}

Figure~\ref{fig:4site_vqs_direct_GF_plots}(a) shows the diagonal component of the Green's function, $G_{1\uparrow,1\uparrow}(\tau)$, computed by VQS or direct VQS as well as the exact one .
The imaginary-time evolution was performed on 70 sampling frequencies in the same manner as for the dimer model.
As $G_{ii}(\tau) = -G_{ii}(-\tau)$ holds owing to the particle--hole symmetry, we only show the data for $\tau>0$.
The exact Green's function decays slowly for $10\lesssim \tau \lesssim 10^2$ owing to the small gap at first and then vanishes exponentially for $\tau \gtrsim 10^2$.

 Figure~\ref{fig:4site_vqs_direct_GF_plots}(a) clearly demonstrates that the Green's functions can be accurately computed using our algorithms.
At $\tau=0$, the computed Green's function agrees with the exact value within an accuracy of $10^{-5}$.
This error comes from the fitting in stage 2.
As $\tau$ increases, the absolute error increases owing to the discretization error in $\tau$.
The relative error however seems to stay constant at large $\tau$,
indicating the numerical stability of the present method.

Figure~\ref{fig:4site_vqs_direct_GF_plots}(b) shows the exponential component $e^{\eta(\tau)+\tau E_\mathrm{G}}$ of Eq.~\eqref{eq:final_form_of_gr} and the remainder computed by VQS.
For $10\lesssim \tau \lesssim 10^2$, the remaining part depends on $\tau$, demonstrating that the slow decay of $G_{1\uparrow,1\uparrow}(\tau)$ is determined not only by the exponential part but also by the transition amplitude. 
For $\tau \gtrsim 10^2$, the remaining part is constant, where $e^{\eta(\tau)+\tau E_\mathrm{G}}$ dominates the exponential decay of $G_{1\uparrow,1\uparrow}(\tau)$ at large $\tau$.

\begin{figure}[h]
    \includegraphics[width=0.70\linewidth]{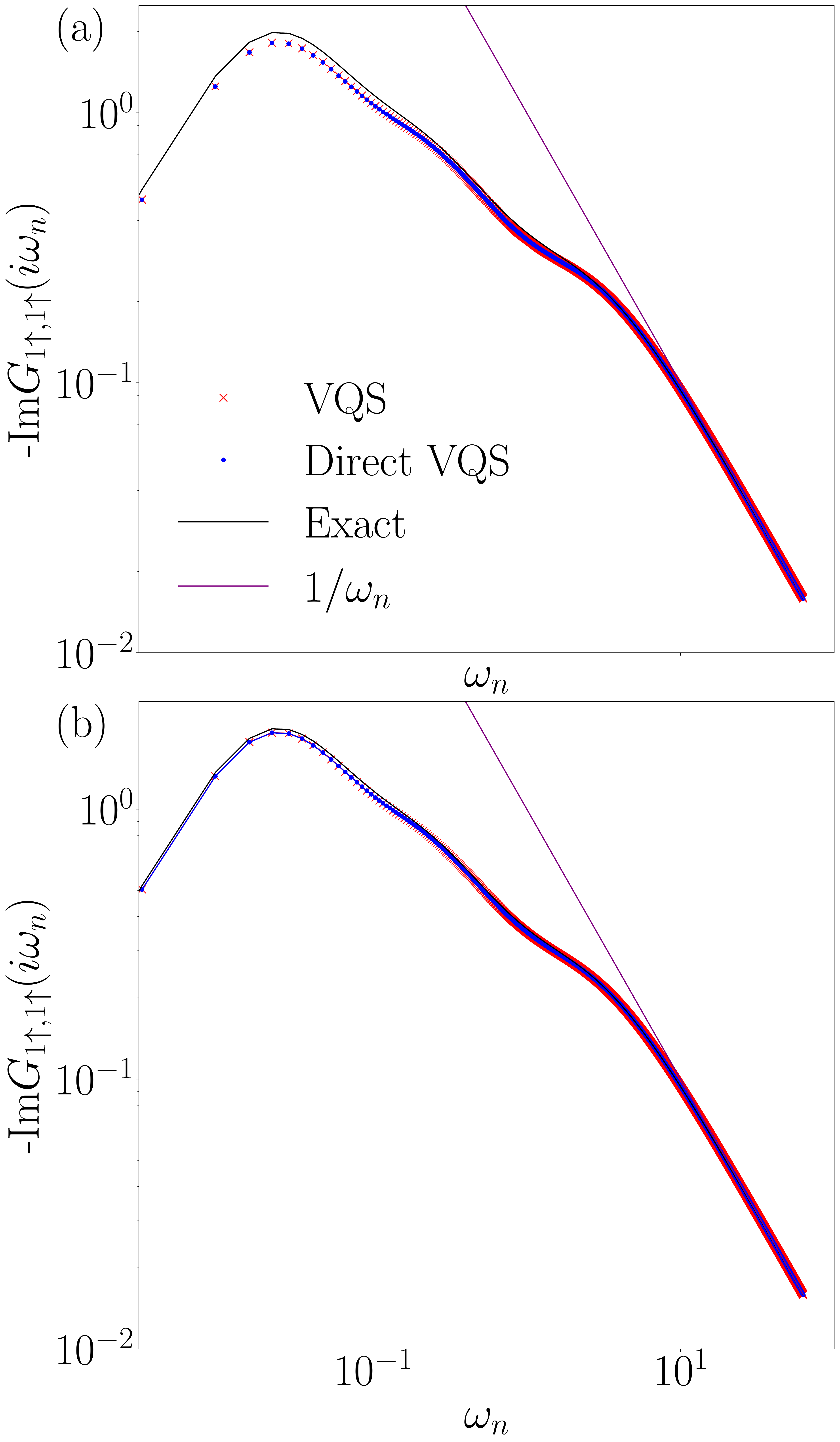}
    \caption{
    Matsubara Green's function computed for the four-site impurity model using 70 (a) and 139 (b) sparse sampling points.
    The data in (a) correspond to the imaginary-time data shown in Fig.~\ref{fig:4site_vqs_direct_GF_plots}.
    }
    \label{fig:4site_vqs_fourier_GF}
\end{figure}

Figure~\ref{fig:4site_vqs_fourier_GF}(a) shows the Matsubara Green's function transformed from the $\tau$ domain.
The Green's functions computed by VQS and direct VQS are in good agreement with the exact function.
The Green's function decays as $1/\ii \omega_n$ for the same reason as the dimer model.
To confirm that this error originates from the discretization error in the imaginary-time evolution, we performed a similar simulation using a finer mesh comprising 139 sampling points, which were constructed by taking the midpoints of the original sampling points.
Figure~\ref{fig:4site_vqs_fourier_GF}(b) shows the result, indicating that the effect is due to the discretization error as the error here is smaller than that shown in Fig.~\ref{fig:4site_vqs_fourier_GF}(a).
Simultaneously, we observed that the error in the imaginary-time Green's function was also reduced (figure is not shown).

\begin{figure}[h]
    \includegraphics[width=1.0\linewidth]{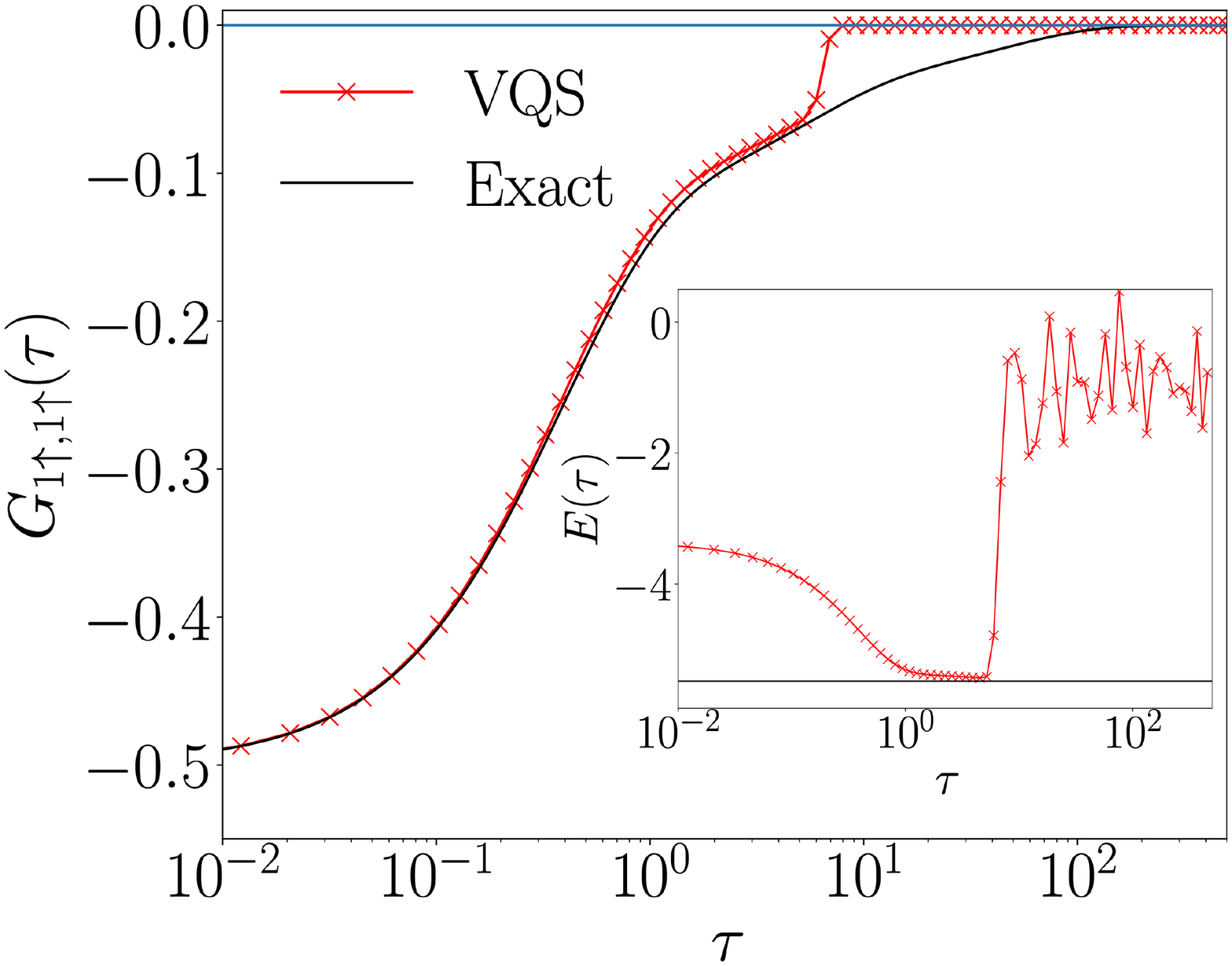}
    \caption{
    Numerical instability in computing the imaginary-time Green's function for a four-site impurity model by VQS ($\tau>0$).
    The panel shows Green's function computed without the adaptive construction of the mesh (see Appendix~\ref{sec:safe_methods}).
    The inset shows $E(\tau)$ and the exact ground-state energy for $N+1$ particles (black horizontal line).
    }
    \label{fig:failed_4site}
\end{figure}

As indicated by the thin vertical lines in Fig.~\ref{fig:4site_vqs_direct_GF_plots}(b),
to avoid numerical instability,
we performed an imaginary-time evolution using an adaptively generated mesh in the $\tau$ domain.
We decreased $\Delta \tau$ when the imaginary-time evolution became unstable (for more details, refer to Appendix~\ref{sec:safe_methods}).
For the 70 and 139 sparse sampling points shown in Fig.~\ref{fig:4site_vqs_fourier_GF},
we used totally 121 and 159 mesh points for the imaginary-time evolution.
Figure~\ref{fig:failed_4site} shows how the imaginary-time evolution fails without the adaptive procedure.
At $\tau\simeq 5$, the imaginary-time evolution becomes unstable, which is signaled by the sudden increase in $E_\tau$.
There are two possible reasons for this.
First, the first-order approximation of the Taylor expansion in Eq.~\eqref{eq:next_theta_vqs} is no longer a good approximation because $\Delta \tau$ is extremely large.
The second possibility is that the parametrization of the UCCGSD ansatz is redundant.
This results in arbitrariness in the gradient $\dot{\theta}_i(\tau)$,
which makes Eq.~\eqref{eq:derivative_parameters} ill-posed.
In other words, the condition number of $M$ diverges.
This may be because our test cases are still limited to small systems owing to the expensive computational cost of the imaginary-time on a quantum circuit simulator.
To investigate this problem, solving a larger system with the same ansatz is necessary.
Another interesting approach may be to construct a more compact ansatz with fewer  parameters.

\subsection{Robustness to shot noise of imaginary-time evolution by VQS}\label{sec:four-site}
Estimating the minimal number of shots for measurements assuming a fault-tolerant quantum computer is important.
A previous study provides an analytic expression on the estimate of the total number of measurements for computing Green's functions by real-time evolution by VQS \cite{mcardle2019variational}.

However, we cannot apply their expression to the computation of $G(\tau)$ in a straightforward way due to the following reasons.
The estimate of the number of shots depends on many factors such as the ansatz, the grouping of the terms of the Hamiltonian in measurements, 
and the condition number of the linear equation in Eq.~\eqref{eq:derivative_parameters} in VQS.
In particular, the condition number of the matrix $M$ in Eq.~\eqref{eq:Aij} also strongly depends on the system and the ansatz used. 
Actually, the condition number is divergent in the present cases.
Thus, we somehow have to resort to numerical simulations.

Instead of estimating the total number of measurements for computing $G(\tau)$ by imaginary-time evolution, 
we performed numerical simulations on the stability of our algorithm against shot noise  assuming a fault-tolerant quantum computer.
To be more specific, we studied how shot noise in the matrix and vector elements at an each time step [Eq.~(28) and Eq.~(29)] affects the imaginary-time evolution by VQS and the computed $G(\tau)$.
The shot noise was emulated by adding Gaussian noise with mean 0 to each element.
The width of the distribution was set to the product of $\sigma$ and the exact value,
where $\sigma~(> 0)$ denotes the relative amplitude of the shot noise.

We performed simulations with various values of $\sigma$ for the dimer model and the four-site model.
In the following simulation,
we removed the methods to deal with the numerical instability 
such as the ``energy convergence condition'' 
and ``additional imaginary-time points'' 
as discussed in Appendix~\ref{sec:safe_methods}.

\subsubsection{Dimer model}
\begin{figure}[h]
    \includegraphics[width=1.0\linewidth]{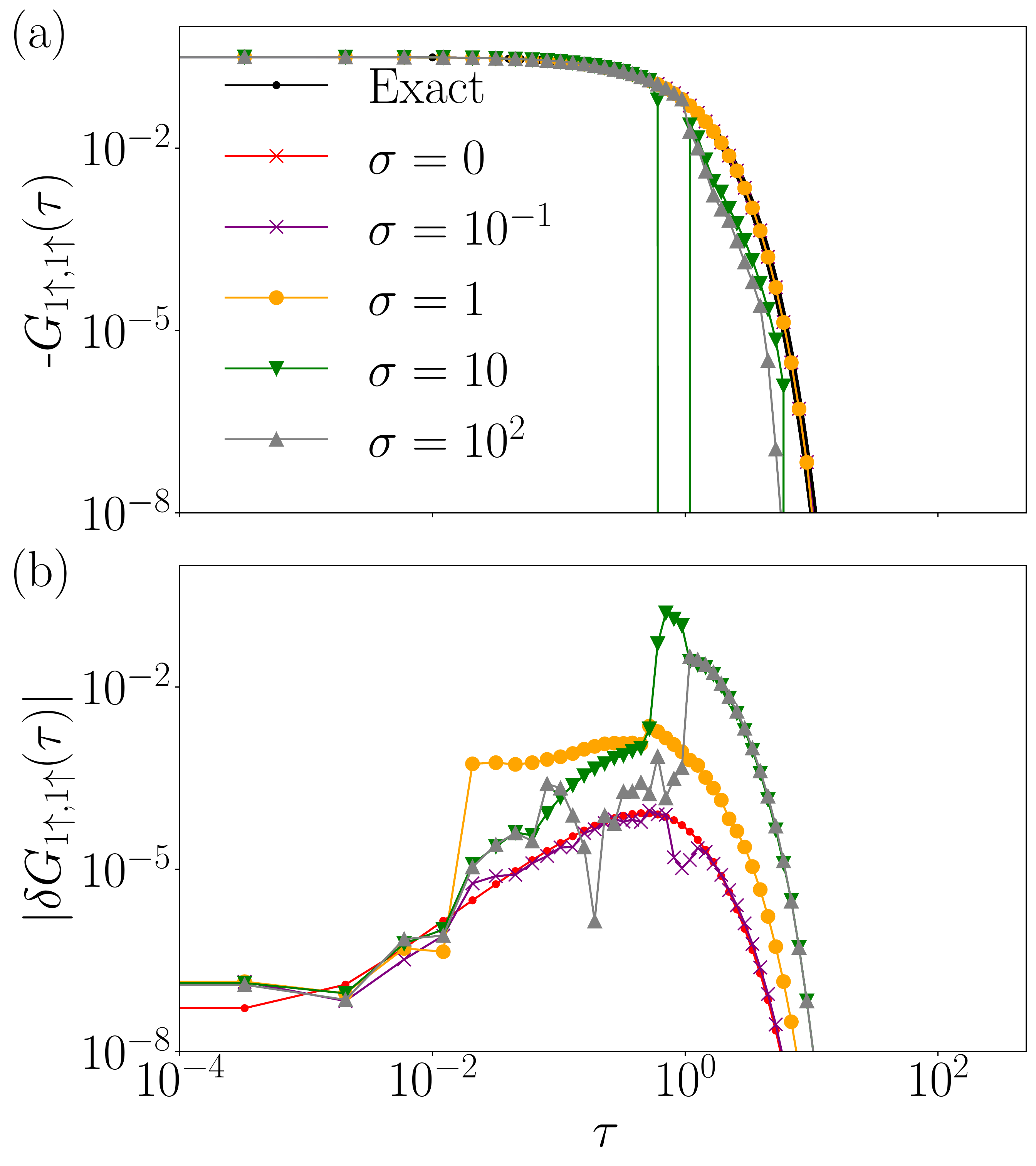}
    \caption{
        Computed $G_{1\uparrow,1\uparrow}(\tau)$ (a) and the error (b)
    at various shot-noise levels
    for the dimer model. 
    }
    \label{fig:dimer_plus_vqs_g_shotnosie}
\end{figure}

Figures~\ref{fig:dimer_plus_vqs_g_shotnosie}(a) and (b) show computed $G_{1\uparrow,1\uparrow}(\tau)$ 
and the error in $G_{1\uparrow,1\uparrow}(\tau)$, respectively.
Remarkably, the imaginary-time evolution is stable up to as large as $\sigma=1$.
As the exponential part of the Green's function decays exponentially at $\tau \gtrsim 10^0$, 
the error also decays exponentially.
For $\sigma \gtrsim 10$, the imaginary-time evolution fails at $\tau \simeq 1$ since the errors $|\delta G_{1\uparrow,1\uparrow}(\tau)|$ in the computed $G_{1\uparrow,1\uparrow}(\tau)$ are accumulated.

For reference, we demonstrate how the same shot noise 
as VQS affected the ground energy obtained by VQE.
The shot noise was emulated by adding Gaussian noise with mean=0 to the expectation value of the Hamiltonian.
When the width of the distribution $\sigma$ = [$10^{-5}$, $10^{-3}$, $10^{-1}$, $1$, $10$, $100$],
the relative errors of the expectation values are [$6.545 \times 10^{-6}$, $0.001$, $0.065$, $0.460$, $3.217$, $169.851$], respectively.
To keep the relative error in energy low enough compared to the energy gap between the ground state and the first excited state, it will be necessary to reduce the shot noise to $\sigma=10^{-3}$.

\subsubsection{Four-site model}
\begin{figure}[h]
    \includegraphics[width=0.9\linewidth]{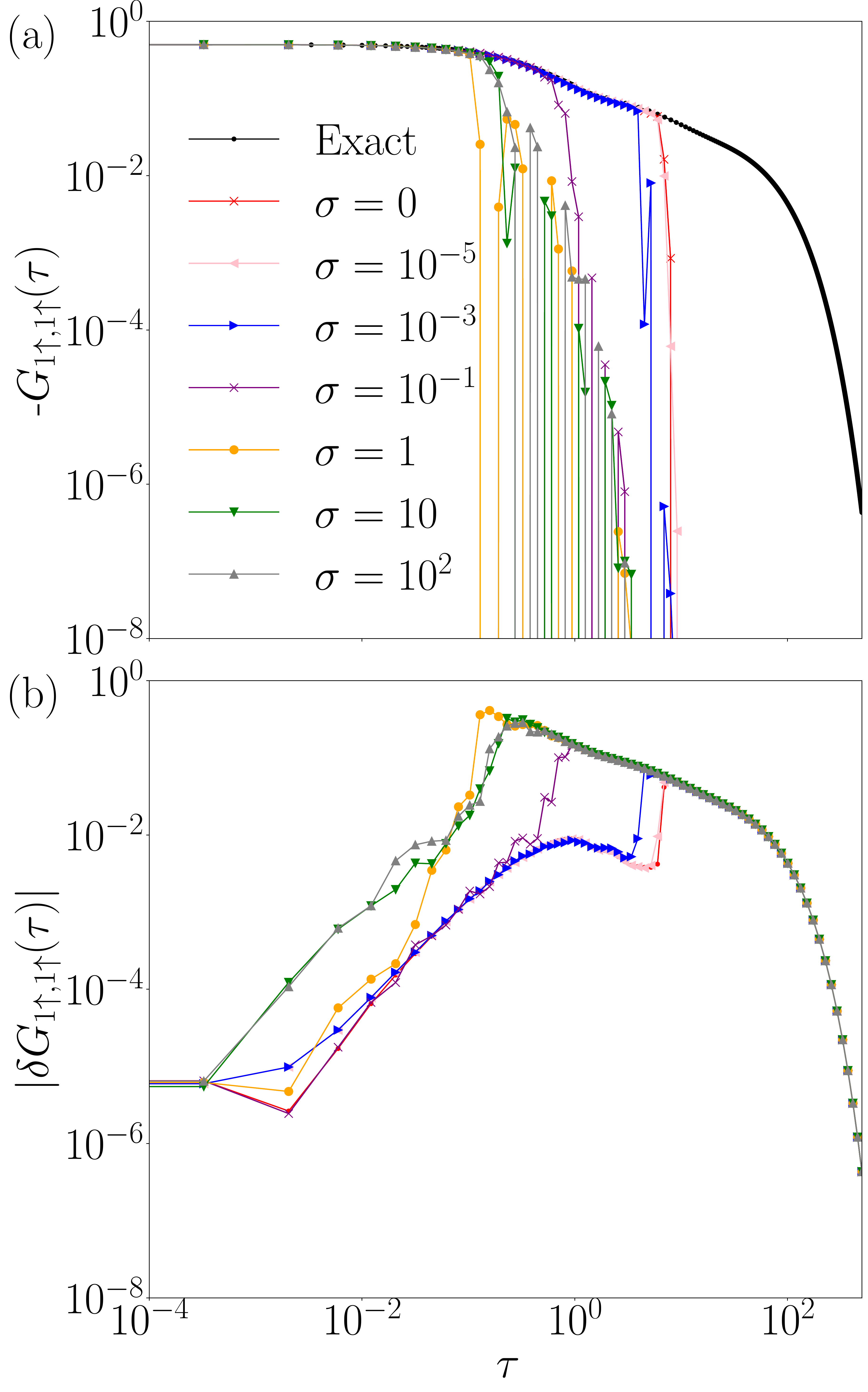}
    \caption{
    Computed $G_{1\uparrow,1\uparrow}(\tau)$ (a) and the error (b)
    at various shot-noise levels
    for the four-site model. 
    }
    \label{fig:4site_plus_vqs_g_shotnosie}
\end{figure}

Figures~\ref{fig:4site_plus_vqs_g_shotnosie}(a) and (b) show the diagonal component of the Green's function $G_{1\uparrow,1\uparrow}(\tau)$ and the error in the computed $G_{1\uparrow,1\uparrow}(\tau)$, respectively.

We confirmed that the imaginary-time evolution is stable up to as large as $\sigma=10^{-3}$.
For $\sigma \gtrsim 10^{-1}$, the imaginary-time evolution becomes more unstable than in the case of $\sigma=0$.

We demonstrated the VQE adding the Gaussian noise in the measurement of the expectation value of the Hamiltonian.
When the energy relative error for the width of the Gaussian distribution 
$\sigma$ = [$10^{-5}$, $10^{-3}$, $10^{-1}$, $1$, $10$, $100$],
the relative errors in energy are [$5.144 \times 10^{-6}$, $0.0004$, $0.032$, $0.5091$, $5.3307$, $65.449$], respectively.
To keep the energy error low enough to compare to the energy gap between the ground state and the first excited state, it will be necessary to reduce the shot noise to $\sigma=10^{-3}$.

\section{Summary AND DISCUSSION}\label{sec:summary}
We proposed a quantum--classical hybrid algorithm to compute the imaginary-time Green's function on quantum devices with limited hardware resources by applying the VQS, 
which has been used to calculate the ground state.
Using the quantum circuit simulator Qulacs, we verified  this algorithm by computing Green's functions for typical impurity models such as the dimer model and four-site impurity model obtained by DMFT.
The imaginary-time Green's function and 
Matsubara Green's function obtained using our algorithm agree well with the exact solution. 
Furthermore, we efficiently computed the imaginary-time Green's function by using a nonuniform mesh to reduce the number of imaginary-time $\tau$ points.
For numerical instabilities occurring in regions where the mesh width $\Delta \tau$ is large, we also computed the Green's function stably by applying a technique of adaptively generating mesh and imposing an energy convergence condition.

Quantum algorithms for computing the Green's function on quantum devices with limited hardware resources have been actively studied in recent years, and the complexity of quantum circuits needs to be discussed.
First, we discuss the scalability of our algorithm and compare it with other similar methods.
In calculating the excited states and VQS in our algorithm, the measurement must be repeated many times in the transition amplitude algorithm.
This requires only a single ancillary qubit, which is approximately twice the depth of the quantum circuit used to calculate the ground state of the VQE.
However, it would require many two-qubit unitary gates, which are challenging to implement in NISQ devices.

Recently, variational quantum algorithms to directly obtain the Green's function in the frequency domain was developed~\cite{chen2021variational, jamet2021krylov}.
These algorithms may be more general in that they directly compute the real-time Green's function as well as the imaginary-frequency Green's function.
Finite-temperature static and dynamic correlation functions of spin systems
have been calculated using the quantum imaginary time evolution (QITE) algorithm on a five-qubit IBM quantum device~\cite{PRXQuantum.2.010317}.
These algorithms require measuring the expectation value of the square of the Hamiltonian or an effective Hamiltonian, which may be computationally demanding for a larger impurity model, whereas our method does not require measuring these observables.
It is an interesting question which method is more efficient and stable in the presence of realistic noise.

Finally, we discuss future directions.
It is interesting to perform simulations under realistic noise conditions with error mitigation techniques~\cite{rogers2021error, endo2018practical} for comparing the efficiency of the recently proposed various methods.
For clarifying the causes of the numerical instability in the imaginary-time evolution, it is desired to apply the proposed methods to larger impurity models using a more compact/efficient ansatz. Possible directions are tensor decomposition methods~\cite{matsuzawa2020jastrow, rubin2022compressing, PRXQuantum.2.040352, peng2017highly, motta2021low, PRXQuantum.2.030305} and an adaptive variational quantum imaginary-time evolution (AVQITE) approach~\cite{gomes2021adaptive, mukherjee2022comparative}.

\begin{acknowledgments}
R.S. and H.S. were supported by JSPS KAKENHI Grants No. 18H01158, No. 21H01041, and No. 21H01003, and JST PRESTO Grant No. JPMJPR2012, Japan.
W.M. was supported by JST PRESTO Grant No. JPMJPR191A and MEXT Quantum Leap Flagship Program (MEXT QLEAP) Grants No.~JPMXS0118067394 and No. JPMXS0120319794.
W.M. also acknowledges the JST COI-NEXT program JPMJPF2014.
The crystal structure in Fig.~\ref{fig:dmft} was drawn using VESTA~\cite{momma2008vesta}.
\end{acknowledgments}

\bibliography{ref}

\appendix 

\section{Measurement circuits of VQS}\label{sec:measurement circuit of vqs}
This appendix reviews quantum circuits that efficiently measure $M$ and $C$ in Eqs.~\eqref{eq:Aij} and \eqref{eq:Ci}~\cite{mcardle2019variational}.

In general, each unitary operator $U_{i}$ depends only on the parameter $\theta_{i}$.
Assuming that $U_{i}(\theta_{i})$ is either a rotational gate or a controlled rotational gate, 
its derivative can be written as
\begin{align}
    \pdv{U_{i}(\theta_{i})}{\theta_{i}}=\sum_{k} f_{k, i} U_{i}(\theta_{i}) P_{k, i},
\end{align}
where $P_{k,i}$ is the unitary operator and $f_{k, i}$ is the coefficient.
The derivative of the variational quantum state $\ket{\phi(\vec{\theta}(\tau))} = U(\vec{\theta})\ket{0}$  is 
\begin{align}
    \pdv{\ket{\phi(\vec{\theta}(\tau))}}{\theta_{i}}=\sum_{k} f_{k, i} \tilde{U}_{k, i}(\vec{\theta})\ket{0},
\end{align}
where 
\begin{align}
\tilde{U}_{k, i}=U_{N_{P}}(\theta_{N_{P}}) \ldots U_{i+1}(\theta_{i+1}) U_{i}(\theta_{i}) P_{k, i} \ldots U_{2}(\theta_{2}) U_{1}(\theta_{1}).
\end{align}

Then, Eqs.~\eqref{eq:Aij} and \eqref{eq:Ci} can be written as 
\begin{align}
        \label{Mij_}
        M_{ij} 
        &=
        \mathcal{R}(\sum_{k,l} f_{k, i}^{*} f_{l, j} \bra{0}\tilde{U}^{\dagger}_{k,i}(\vec \theta(\tau))
        \tilde{U}_{l,j}(\vec \theta(\tau))\ket{0}),
        \\
        \label{Ci_}
        C_i 
        &=
        - \mathcal{R} (\sum_{k,l}f_{l, j}h_{k}\bra{0}U^{\dagger}(\vec \theta(\tau)) S_{k}
          \tilde{U}_{l,j}(\vec \theta(\tau))\ket{0}
         ).
    \end{align}

\begin{figure}[h]
    \includegraphics[width=1.0\linewidth]{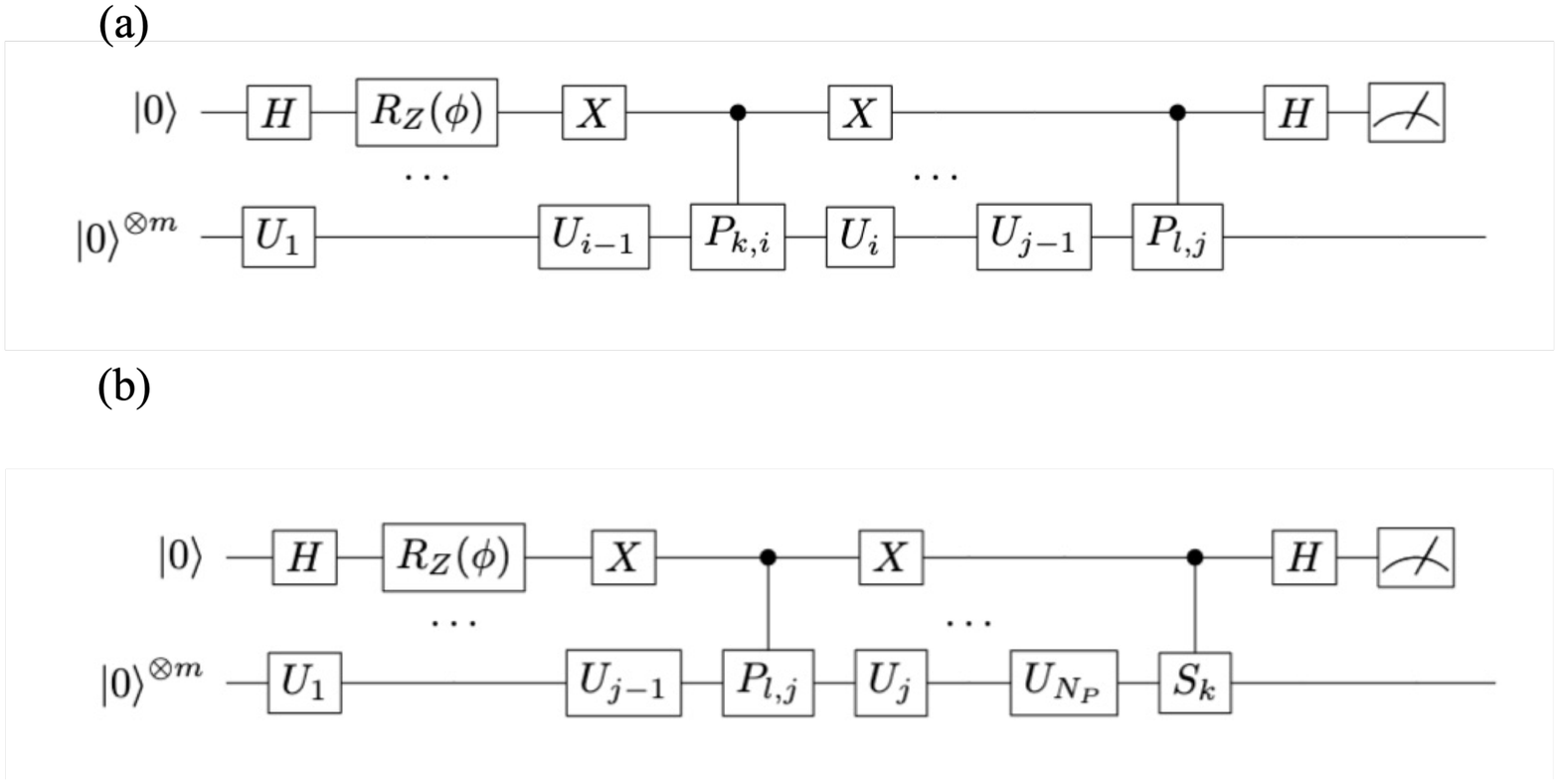}
    \caption{Quantum circuits for measuring the elements of $M$ and $C$.
    (a) $\mathcal{R}(\bra{0}\tilde{U}^{\dagger}_{k,i}(\vec \theta(\tau))
    \tilde{U}_{l,j}(\vec \theta(\tau))\ket{0})$, 
    (b) $\mathcal{R}(\bra{0}U^{\dagger}(\vec \theta(\tau)) S_{k}
    \tilde{U}_{l,j}(\vec \theta(\tau))\ket{0})$.
    }
    \label{fig:measurement_circuits_vqs}
\end{figure}

Assuming $i<j$, the component of Eq.~\eqref{Mij_} can be computed as follows:
\begin{multline}
\bra{0}\tilde{U}^{\dagger}_{k,i}(\vec \theta(\tau))
    \tilde{U}_{l,j}(\vec \theta(\tau))\ket{0} \\
    =\bra{0}U_{1}^{\dagger}(\theta_{1}) \ldots U_{i-1}^{\dagger}(\theta_{i-1})P^{\dagger}_{k, i} U^{\dagger}_{i}(\theta_{i})\ldots U^{\dagger}_{N_{P}}(\theta_{N_{P}}) \\
    U_{N_{P}}(\theta_{N_{P}}) \ldots U_{j}(\theta_{j}) P_{l, j} U_{j-1}(\theta_{j-1})\ldots  U_{i}(\theta_{i}) \ldots  U_{1}(\theta_{1})
    \ket{0} \\
    =\bra{0}U_{1}^{\dagger}(\theta_{1}) \ldots U_{i-1}^{\dagger}(\theta_{i-1})P^{\dagger}_{k, i} U^{\dagger}_{i}(\theta_{i})\ldots U^{\dagger}_{j-1}(\theta_{j-1}) \\
     P_{l, j} U_{j-1}(\theta_{j-1})\ldots  U_{i}(\theta_{i}) \ldots  U_{1}(\theta_{1})
    \ket{0},\\
\end{multline}
where $\mathcal{R}( f_{k, i}^{*} f_{l, j} \bra{0}\tilde{U}^{\dagger}_{k,i}(\vec \theta(\tau))
\tilde{U}_{l,j}(\vec \theta(\tau))\ket{0})$
can be measured by the quantum circuit in Fig.~\ref{fig:measurement_circuits_vqs}(a).

We can compute the component of Eq.~\eqref{Ci_} in the same way: 
\begin{multline}
    \bra{0}U^{\dagger}(\vec \theta(\tau)) S_{k}
    \tilde{U}_{l,j}(\vec \theta(\tau))\ket{0} \\
        =\bra{0}U_{1}^{\dagger}(\theta_{1}) \ldots U_{N_{P}}^{\dagger}(\theta_{N_{P}}) S_{k} \\
        U_{N_{P}}(\theta_{N_{P}}) \ldots U_{j}(\theta_{j}) P_{l, j} U_{j-1}(\theta_{j-1})\ldots  U_{i}(\theta_{i}) \ldots  U_{1}(\theta_{1})
        \ket{0},
\end{multline}
where  $\mathcal{R}(\bra{0}U^{\dagger}(\vec \theta(\tau)) S_{k}
\tilde{U}_{l,j}(\vec \theta(\tau))\ket{0})$
can be measured by the quantum circuit in Fig.~\ref{fig:measurement_circuits_vqs}(b).

\section{Ansatz for Variational Quantum Algorithms}\label{sec:details for vqa}

As a parametric quantum circuit (i.e., ansatz) for VQE and VQS, we choose the unitary coupled cluster with generalized singles and doubles (UCCGSD)~\cite{lee2018generalized,nooijen2000can}, 
which is an extension of coupled cluster methods widely used in the fields of quantum chemistry.
In general, one can use different ansatze for the ground state in stage 1 and the excited states in stage 2 and stage 3.
For simplicity, however, we use the same ansatz from stage 1 to stage 3 in the present study.

The UCCGSD is defined in the following form:
\begin{align}
   \ket{\Psi_{\mathrm{U C C G S D}}} &=e^{\hat{E}+\hat{D}}\ket{\Psi_{\mathrm{init}}},
\end{align}
where $\ket{\Psi_{\mathrm{init}}}$ is a product state, and
$\hat{E}=\hat{T_{1}}-\hat{T_{1}}^{\dagger}$ and $\hat{D}=\hat{T_{2}}-\hat{T_{2}}^{\dagger}$ consist of excitation operators $\hat T_{n}$ $(n=1,2)$ and their conjugates.
The excitation operators $\hat{T_n}$ are

\begin{align}
    \hat{T_{1}} =\sum_{p q} t_{q}^{p} \hat c_{q}^{\dagger} \hat c_{p},
\end{align}

\begin{align}
    \hat{T_{2}} = \frac{1}{4} \sum_{p q, r s} t_{r s}^{p q} \hat c_{p}^{\dagger} \hat c_{q}^{\dagger} \hat c_{s} \hat c_{r},
\end{align}

\noindent
where $\hat{T_1}$ is the single-particle excitation operator, and $\hat{T_2}$ is two-particle excitations. 
The indices $p, q, r, s$ run over all the spin orbitals in UCCGSD, which is a generalization of unitary coupled cluster (UCC) \cite{kutzelniggQuantumChemistryFock1982,kutzelniggQuantumChemistryFock1983,kutzelniggQuantumChemistryFock1985,bartlettAlternativeCoupledclusterAnsatze1989,kutzelniggErrorAnalysisImprovements1991,taubeNewPerspectivesUnitary2006}  with respect to the sum index of excitation operators. 
Thus, the UCCGSD includes single-particle basis rotations in the spin-orbital space because $e^{\hat  E}\ket{\Psi_{\mathrm{init}}} $ in Eq.~\eqref{eq:uccgsd_trotter} is an orbital rotation unitary operator ~\cite{mizukami2020orbital}.

The UCCGSD is not efficiently computable on a classical computer because we need to compute the each term of the expansion of $\expval{e^{{\hat{E^{\dagger}}+\hat{D^{\dagger}}}} H e^{\hat{E}+\hat{D}}}{0}$, which continues infinitely. 
On the other hand, on a quantum computer, we can compute the expectation value directly.

Since terms in $\hat{E}+\hat{D}$ do not commute with each other, the implementation of UCCGSD on a quantum computer requires us to approximate the exponential $e^{\hat{E}+\hat{D}}$ by a Trotter decomposition.
However, it is known that this Trotter error can be largely absorbed in the process of classical optimization with the flexibility of the variational quantum algorithm~\cite{o2016scalable,barkoutsos2018quantum}.

Therefore, we set the Trotter step to 1 as
\begin{align}
   \label{eq:uccgsd_trotter}
   &\ket{\Psi_{\mathrm{UCCGSD}}}\simeq e^{\hat D}e^{\hat E}\ket{\Psi_{\mathrm{init}}} \\
   \label{eq:uccgsd_trotter2}
   &=\prod^{n_\mathrm{so}}_{p,q,r,s} 
   \{e^{t^{pq}_{rs} \hat c^\dagger_p \hat c^\dagger_q c_s c_r - t^{pq*}_{rs} \hat c^\dagger_r \hat c^\dagger_s \hat c_q \hat c_p}\}\times\nonumber\\
   &\hspace{3em}\prod^{n_{\mathrm{so}}}_{p,q}  
   \{e^{t^{p}_{q} \hat c^\dagger_p \hat c_q -t^{p*}_{q} \hat c^\dagger_q \hat c_p }\}
   \ket{\Psi_{\mathrm{init}}}, 
\end{align}
where $\ket{\Psi_{\mathrm{init}}}$ has the same number of particles as spatial orbitals and $n_{\mathrm{so}}$ is all spin orbitals.
$\hat T^{p q}_{r s}$ and $\hat T^{p}_{q}$ are replaced by real variational parameters on a quantum computer in performing VQE and VQS.

$e^{\hat E}\ket{\Psi_{\mathrm{init}}}$ in Eq.~\eqref{eq:uccgsd_trotter} includes a rotation in the spin-orbital space~\cite{mizukami2020orbital}.
This implies that we get optimal orbitals for a correlated wave function via optimizing orbital rotation parameters on a classical computer.

Because the UCCGSD conserves the number of particles, the variational quantum state $\ket{\Psi_{\mathrm{UCCGSD}}}$ has the same number of particles as the initial state $\ket{\Psi_{\mathrm{init}}}$.
In contrast, the total $\hat S^{z}$ is not necessary conserved in UCCGSD.
In this study, we fix the total $\hat S^{z}$ of the variational quantum state by removing the operators $\hat T_{i}^{a}$ and $\hat T_{i j}^{a b}$ that mix different spins.

\section{Sparse mesh and Fourier transform}\label{sec:ft}
The present method of the computation of the imaginary-time Green's function can be combined with an arbitrary mesh in the imaginary-time space.
Nevertheless, the use of an appropriate non-uniform mesh significantly reduces discretization errors because the Green's function varies rapidly only in the vicinity of $\tau= n\beta$ ($n=0,\pm 1, \pm 2$).
In the numerical simulations shown in the following sections, we adopt a sparse mesh generated according to a compact orthogonal basis of $G(\tau)$~\cite{Li:2020eu}, the intermediate-representation (IR) basis~\cite{Shinaoka:2017ix,Chikano:2018gd}.
The sparse mesh is dense near $\tau=0$ and becomes sparse around $\tau=\beta/2$ (see Fig.~\ref{fig:dimer_vqs_direct_GF_plots}).
Numerical data of $G(\tau)$ on the sparse mesh points can be transformed 
to Matsubara frequencies through the IR basis without discretizing Eq.~\eqref{eq:matsubara}~\cite{Li:2020eu}.
We refer the interested reader to a recent review article~\cite{shinaoka2021efficient}.

\section{Improving numerical instability of imaginary-time evolution}\label{sec:safe_methods}
As we demonstrated in Sec.~\ref{sec:four-site},
the imaginary-time evolution in stage 3 sometimes becomes unstable especially when the time step $\Delta\tau$ is large.
In stage 3, the energy $E_\tau$ must decease.
We however observed that $E_\tau$ starts to rise at a certain imaginary time possibly due to numerical instability.

To improve this numerical stability further, we introduce the following additional tricks.
First, if $E_\tau$ is increased in imaginary-time evolution from $\tau$ to $\tau+\Delta \tau$,
we first perform time evolution with a smaller time step from $\tau$ to $\tau + \Delta \tau/2$, which is followed by time evolution from $\tau+\Delta \tau/2$ to $\tau + \Delta \tau$.
We apply this strategy recursively to these two sections until the numerical instability is removed.
Second, once $E_\tau$ converges, i.e., if $\frac{|E(\tau_{N})-E(\tau_{N+1})|}{\Delta \tau}<10^{-5}$ is met,
we stop the imaginary-time evolution of the variational parameters. 

To avoid this problem, 
as proposed in Ref.~\onlinecite{mcardle2019variational}, we truncate the singular values smaller than $10^{-5}$ multiplied by the maximum singular value
in solving the linear equation of Eq.~\eqref{eq:derivative_parameters}.

\end{document}